\documentclass[12pt]{iopart}

\usepackage{psfrag,graphicx}
\usepackage{dcolumn}
\usepackage{amsmath,amssymb}
\usepackage{bm}
\usepackage{amsfonts,amssymb,amsmath}        

\newcommand{\be}{\begin{equation}}
\newcommand{\ee}{\end{equation}}
\newcommand{\bq}{\begin{eqnarray}}
\newcommand{\eq}{\end{eqnarray}}

\newcommand{\Alpha}{\mbox{\boldmath${\alpha}$}}

\newcommand{\ket}[1]{\left | \, #1 \right\rangle}

\newcommand{\rf}[1]{(\ref{#1})}
\begin{document}

\title{Interacting non-Abelian anyons as Majorana fermions in the honeycomb lattice model }

\author{Ville Lahtinen}
\address{NORDITA, Roslagstullsbacken 23, 106 91, Stockholm, Sweden}
\date{\today}
\ead{ville.lahtinen@nordita.org}

\begin{abstract}

We study the collective states of interacting non-Abelian anyons that emerge in Kitaev's honeycomb lattice model. Vortex-vortex interactions are shown to lead to the lifting of the topological degeneracy and the energy is discovered to exhibit oscillations that are consistent with Majorana fermions being localized at vortex cores. We show how to construct states corresponding to the fusion channel degrees of freedom and obtain the energy gaps characterizing the stability of the topological low energy spectrum. To study the collective behavior of many vortices, we introduce an effective lattice model of Majorana fermions. We find necessary conditions for it to approximate the spectrum of the honeycomb lattice model and show that bi-partite interactions are responsible for the degeneracy lifting also in many vortex systems.

\end{abstract}

\pacs{05.30.Pr, 74.25.Uv, 75.10.Jm}

\maketitle

\section{Introduction}

One of the main trends in contemporary condensed matter physics is the study of topological order. Apart from their fundamental interest, the motivation derives from the discovery of topological quantum computation \cite{Kitaev03,Freedman04}. The quasiparticle excitations of topologically ordered systems are anyons that can be employed to perform fault resilient quantum information processing \cite{Nayak, Brennen}. It is of interest to find out what are the sufficient conditions for topological order to exist, which systems are useful for quantum computation and how anyons emerge. While the research concentrated initially on the experimentally challenging fractional quantum Hall states or abstract spin lattice constructions, the recently discovered topological insulators have given rise to novel constructions of anyon supporting systems \cite{Hasan10}. The experimental accessability of the latter has led to the theory and detection of Majorana fermions becoming the subject of intense research.

Majorana fermions are the simplest possible non-Abelian anyons \cite{Rowell09}. While they are not universal for quantum computation \cite{Bravyi05, Bonderson10}, their realization and the subsequent demonstration of topological information processing is considered a significant stepping stone towards a full scale implementation. Physical systems that are predicted to give rise to Majorana fermions include $p$-wave superconductors \cite{Read00}, Kitaev's honeycomb lattice model \cite{Kitaev05} and its generalizations \cite{Yao07, Kells10-2,Yang07,Yao09,deGottardi11}, the Moore-Read state in fractional quantum Hall systems \cite{Read92} and lately also various superconductor/insulator interfaces \cite{Fu08,Sau10, Wimmer10}. Experiments have been performed on fractional quantum Hall states \cite{Willett09} as well as on topological insulator interfaces \cite{Hasan10}, but the existence of Majorana fermions remains still ambiguous. For the detection of Majorana fermions, and for their future applications, it is crucial to understand how their anyonic properties are manifest in the microscopic physics. The essential property to all topological quantum computing schemes is the existence of non-local topological degrees of freedom. For Majorana fermions they appear as fusion degrees of freedom that describe the different locally indistinguishable ways the anyons can behave collectively. However, these states can become locally distinguishable due to tunneling processes that lead to the degeneracy being lifted \cite{Bonderson09}. This has been studied in Kitaev's honeycomb lattice model \cite{Lahtinen08}, in the Moore-Read state \cite{Baraban09} as well as in $p$-wave superconductors \cite{Cheng09, Cheng10, Mizushima10}. It has also recently been discovered that in many anyon systems the tunneling processes can not only lift the degeneracy, but can even drive transitions between topological phases \cite{Feiguin07,Gils09,Ludwig11}. The physical conditions under which this occurs, however, depend on the microscopics of a specific model. Only when the tunneling is exponentially suppressed, the topological phase is stable and its anyonic excitations can be employed for topologically protected quantum computing. 

In the first part of the paper we analyze the degeneracy lifting in the non-Abelian phase of Kitaev's honeycomb lattice model \cite{Kitaev05}. By simulating continuous vortex transport \cite{Lahtinen09}, we extend the vortex-vortex interaction analysis of \cite{Lahtinen08}. We find the anticipated oscillations in the energy splitting \cite{Cheng09}, show how they depend on the microscopics of the model and characterize the stability of the topological low energy spectrum by obtaining the relevant energy gaps. In the second part we introduce an effective Majorana fermion lattice model, an extension of the tight-binding model considered in \cite{Grosfeld}, to study the degeneracy lifting in many vortex systems. Simultaneous interactions between vortices lead in general to a hybridization of the states corresponding to fusion degrees of freedom \cite{Lahtinen10}. We find necessary conditions, i.e. relevant effective flux sectors, for the Majorana model to approximate this subspectrum. The flux sectors are interpreted to arise from the vortices of the underlying model, which provides a direct connection between the microscopic and the effective models. Furthermore, we find that the effective tunneling amplitudes are well approximated by the energy splitting due to vortex-vortex interactions. This confirms that the picture of freely tunneling Majorana fermions applies also in many vortex systems.

This paper is organized as follows. In Section II we review the solution of the honeycomb lattice model using Majorana fermionization. We show how it gives rise to an equivalence between coupling and vortex configurations and how this can be employed to simulate continuous vortex transport. In Section III we analyze the oscillating degeneracy lifting of the fusion degrees of freedom when the separation between two vortices is varied. We obtain the physical parameter dependence of both the oscillations and the energy gaps characterizing the topological low energy spectrum. In Section IV we introduce a lattice model of Majorana fermions and employ it to study the collective behavior of up to eight vortices.

\section{The honeycomb lattice model}

In this section we briefly review Kitaev's honeycomb lattice model and its solution by mapping it to free Majorana fermions. For a more rigorous treatment we refer to the original work \cite{Kitaev05} and to the subsequent developments \cite{Lahtinen08, Pachos07,Kells08,Yu08, Kells09,Vidal08,Chen08}. We show how to relate the manipulation of the vortices to the manipulation of the model's physical parameters. This enables the simulation of continuous vortex transport \cite{Lahtinen09}, which is later used to study the physics of the anyonic vortices.

\subsection{Solution by mapping to free Majorana fermions}

\begin{figure}[t]
\begin{center}
\includegraphics*[width=10cm]{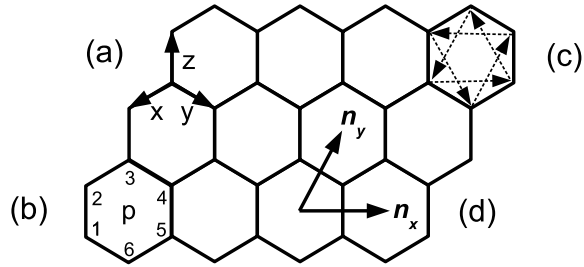}
\end{center}
\caption{\label{honeycomb}The conventions for the honeycomb lattice model. (a) The links are oriented as shown and labeled $x$, $y$ or
$z$ depending on their orientation. (b) A single plaquette $p$ with its six sites enumerated. (c) The oriented six next to nearest neighbour couplings per plaquette originating from the three spin term in the Hamiltonian \cite{Kitaev05}. (d) The normalized lattice basis vectors $\mathbf{n}_x$ and $\mathbf{n}_y$.}
\end{figure}

Kitaev's model \cite{Kitaev05}, comprises of spin-$1/2$ particles residing
on the vertices of a honeycomb lattice. The spins interact according to the
Hamiltonian
\begin{eqnarray} \label{H}
    H = - \sum_{\alpha = x,y,z} \sum_{( i,j) \in \alpha-links} J^\alpha_{ij} \sigma^\alpha_i \sigma^\alpha_j -K \sum_{( i,j,k)} \sigma^x_i \sigma^y_k \sigma^z_j,
\end{eqnarray}
where $J^\alpha_{ij}$ are real nearest neighbour couplings on links of type $\alpha$ specified by their orientation (see Figure
\ref{honeycomb}(a)). The second term is an effective magnetic field
of magnitude $K$, which explicitly breaks time-reversal invariance. The sum in this term runs over all next to nearest
neighbour triplets as described in \cite{Lahtinen08}.

The model can be mapped to a free fermion problem when the spins are represented by Majorana fermions \cite{Kitaev05}. This procedure reduces the Hamiltonian to the quadratic form
\be \label{h}
	H = \frac{i}{4} \sum_{i,j} \hat{h}_{ij} c_{i} c_{j}, \qquad \hat{h}_{ij} = 2J_{ij} \hat{u}_{ij} + 2 K \sum_{k} \hat{u}_{ik} \hat{u}_{jk}.
\ee
Here $c_i = c_i^\dagger$ are Majorana fermions  that satisfy $\{ c_i, c_j\}=2\delta_{ij}$. The first term describes their hopping between nearest neighbour sites while the second term describes next to nearest neighbour hopping between sites shown in Figure \ref{honeycomb}(c). The Hermitian operators $\hat{u}_{ij}=\hat{u}_{ij}^\dagger$ live on the links of the honeycomb lattice and should be understood as a $Z_2$ gauge field. Since $[\hat{u}_{ij},H]=0$, they have no dynamics. The gauge field interpretation follows from the local constraint $D_i \ket{\Psi} = \ket{\Psi}$ that the eigenstates $\ket{\Psi}$ of the original Hamiltonian \rf{H} have to satisfy for all sites $i$ \cite{Kitaev05}. The operator $D_i$ anti-commutes with the three link operators acting on vertex $i$ and thus it can be interpreted as a local gauge transformation. The Hamiltonian \rf{h} is gauge invariant.

At the heart of the exact solvability of the model are the local symmetries $[\hat{w}_p,H]=0$, where, as illustrated in Figure \ref{honeycomb}(b), $\hat{w}_p = \sigma_1^x \sigma_2^y \sigma^z_3 \sigma_4^x \sigma_5^y \sigma_6^z$ are Hermitian plaquette operators associated with every plaquette $p$. In the fermionized picture these become
\be \label{w}
	\hat{w}_p = \hat{u}_{12}\hat{u}_{32}\hat{u}_{34}\hat{u}_{54}\hat{u}_{56}\hat{u}_{16}, \qquad [\hat{w}_p, D_i] = 0,
\ee
that in the gauge theory language can be understood as gauge invariant Wilson loop operators. Their eigenvalues $w_p = - 1$ are interpreted as having a $\pi$-flux {\em vortex} on plaquette $p$. The pattern of all plaquette operator eigenvalues $w \equiv \{ w_p \}$ is referred to as a particular \emph{vortex sector}, that is created by fixing the \emph{gauge}, i.e. choosing the pattern of the link operator eigenvalues $u \equiv \{ u_{ij} \}$. The vortex sectors label the physical sectors of the model.

We consider finite systems of $2 L_x L_y$ spins with $L_x$ and $L_y$ plaquettes in directions $\mathbf{n}_x$ and $\mathbf{n}_y$ (see Figure \ref{honeycomb}(d)).  Due to particle-hole symmetry in the problem, diagonalization of \rf{h} in vortex sector $w$ yields in general the double spectrum
\be \label{Hw}
 H^w = \frac{1}{2} \sum_{i=1}^{L_x L_y} E_i^w \left( b^\dagger_i b_i - b_i b_i^\dagger \right),
\ee 
where $b_i = \sum_{j=1}^{2 L_x L_y} [\Alpha^w_{i}]_j c_j$, for some complex coefficients $[\Alpha^w_{i}]_j$, are complex fermionic modes. When the diagonalization is performed numerically, $\pm E_i^w$ and $\Alpha_i^w$ follow from solving the eigenvalue problem $i h^w \Alpha_i^w = -E_i^w \Alpha_i^w$. Here $h^w$ is the real kernel matrix \rf{h} when restricted to vortex sector $w$.

\subsection{Gauge/coupling configuration equivalence and vortex transport} 

Our aim is to study the physics of the vortices by considering the spectral evolution as a function of the vortex sectors $w$. While the sectors form a discrete set, we can effectively interpolate between them, i.e. simulate the transport the vortices, as follows. As can be seen from \rf{h}, the local value $u_{ij}$ of the gauge field appears always uniquely paired with a local coupling $J_{ij}$. This implies that while the vortex sector depends on the gauge through the plaquette operators, \rf{w}, from the point of view of the Hamiltonian, \rf{h}, we can regard $u$ as the sign configuration of the couplings. It follows that the Hamiltonians $H^w$ and $H^{w'}$ on two different vortex sectors can always be connected by flipping the signs of some of the couplings, i.e. one can always find some coupling configurations $J$ and $J'$ such that $H^w(J)=H^{w'}(J')$ \footnote{Formally one should also manipulate the local values of $K$. However, when $K$ is small compared to the couplings $J_{ij}$, the magnitude of the three spin term does not significantly affect the physics of the vortices and thus controlling the couplings $J_{ij}$ is sufficient. }. Therefore, by manipulating the local couplings we can interpolate the spectrum continuously between different vortex sectors and thereby study the spectral evolution when vortices are created, transported and annihilated. When this is performed adiabatically, the vortex sector $w$ does not strictly speaking change, but in our simulation the spectrum will evolve as if it had. This same method has been employed to evaluate also the non-Abelian statistics of the vortices \cite{Lahtinen09}.

\begin{figure}[t]
\begin{center}
\begin{tabular}{cccc}
\includegraphics*[width=3.5cm]{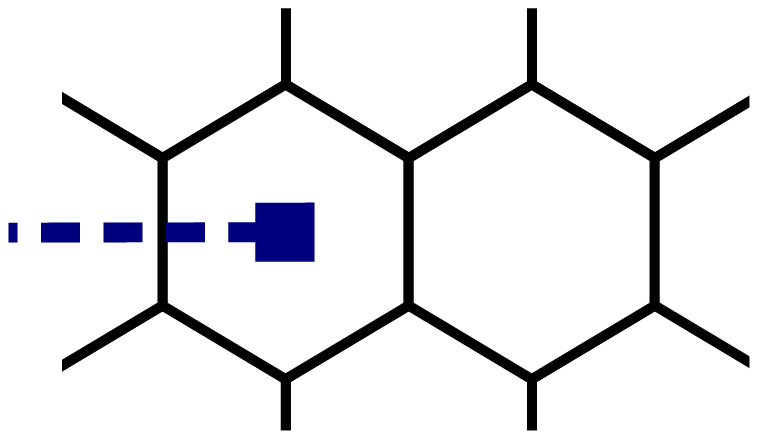} & \includegraphics*[width=3.5cm]{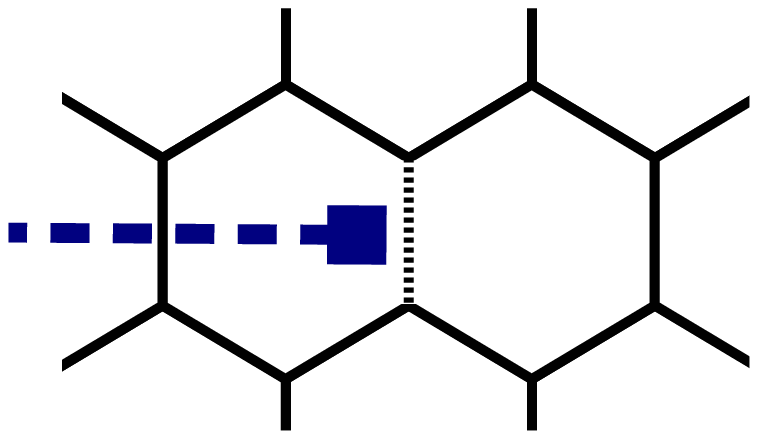} & \includegraphics*[width=3.5cm]{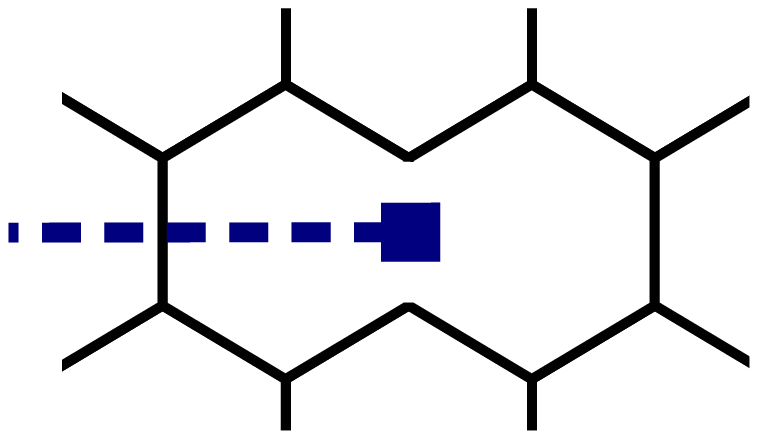} & \includegraphics*[width=3.5cm]{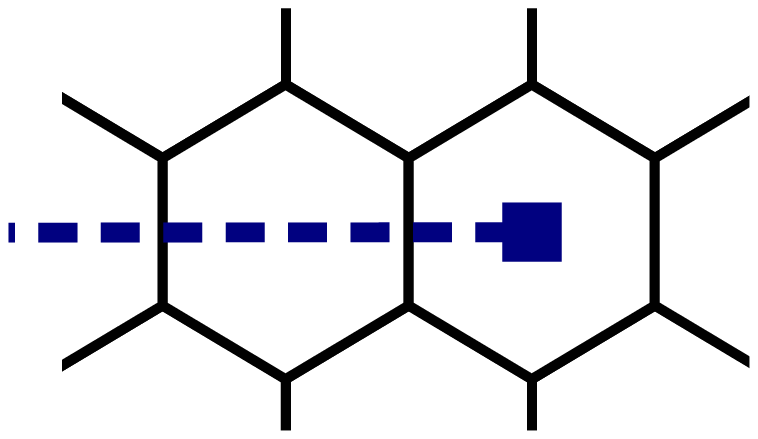} \\
(a) & (b) & (c) & (d)
\end{tabular}
\end{center}
\caption{\label{transport}A protocol for simulating continuous vortex transport. (a) Initially the coupling configuration is chosen such that $J_{ij}=-1$ on the links crossed by the dashed line, while $J_{ij}=1$ on all other links. This corresponds to a vortex on the left plaquette. (b) Consider changing the coupling on the link in the middle from $J_{ij}=1$ to $J_{ij}=-1$ in $S$ steps of size $\frac{2}{S}$. At step $s$ its value is $J_{ij}^s=1-\frac{2s}{S}$, which we interpret as the vortex occupying a location away from the plaquette center. (c) When $J_{ij}^s=0$, the Wilson loop operator is defined only on the composite plaquette, and the vortex is interpreted being located in the middle. (d) Finally, as $J_{ij}^s \to -1$, the vortex moves smoothly to the plaquette on the right.}
\end{figure}

From now on we adopt a perspective where the system has been initially prepared in the ground state belonging to the vortex-free sector \cite{Lieb}. The vortex sectors $w$ are viewed as coupling configurations $J \equiv \{ J_{ij} \}$ of equal amplitude, but with locally varying signs. To transport a vortex between two plaquettes or to create/annihilate a pair of vortices on them, one needs then to change the sign of the physical coupling $J_{ij}$ on the shared link. This can be performed continuously by tuning $J_{ij} \to 0 \to -J_{ij}$ \footnote{Introducing a complex phase is not possible, because the couplings have to be real at all times.}, which we simulate by changing the sign of the coupling in $S$ steps such that at step $s$ its value is given by $J_{ij}(1-\frac{2s}{S})$. As $S$ becomes large, this approximates well a continuous process. The resulting transport protocol is illustrated in Figures \ref{transport}(a)-(d). Treating the vortex sectors $w$ and the coupling configurations $J$ on equal footing is also experimentally motivated. Given sufficient site addressability, the local control of the couplings through external laser fields is how one could perform vortex creation and adiabatic transport in the proposed optical lattice experiments \cite{Duan03, Micheli05}.

\subsection{System of interest}

To study the physics of the anyonic vortices, we employ a large finite toroidal system of 1200 spins defined by $L_x=30$ and $L_y=20$ and consider the system initially in the vortex-free sector by setting $u_{ij}=1$ on all links (any other gauge giving rise to $w_p=1$ on all plaquettes would also do). While this sector of the honeycomb lattice model supports several topological phases where the vortices can behave as either Abelian or non-Abelian anyons \cite{Kitaev05}, we concentrate here only on the latter. It is characterized by Chern number $\nu=\pm1$, which implies that the vortices obey Ising anyon statistics. This phase emerges when $K \neq 0$ and the couplings violate the inequalities $J_\alpha > J_\beta + J_\gamma$ for all all permutations of $\alpha, \beta, \gamma = x,y,z$. Unless otherwise noted, we achieve this by setting $J \equiv \frac{J_x}{J_z}=\frac{J_y}{J_z}=1$ on all links. This amounts to considering the system in the middle of the non-Abelian phase in the sense that the spectral gap is maximized as a function of the couplings $J_\alpha$. For $J<1$ the system is closer to the phase boundary located at $J=\frac{1}{2}$. On top of this background we then imprint different vortex configurations by locally changing the couplings as discussed above. 

We retain the magnitude $K$ of the three spin term as a free parameter. Its physical range depends on the way it is introduced in the proposed optical lattice experiments \cite{Duan03, Micheli05}. If it is introduced by applying a weak Zeeman field \cite{Kitaev05}, only magnitudes of up to $K \approx 10^{-4}$ can be tolerated before the topological phase breaks down \cite{Jiang11}. On the other hand, by engineering it directly one can in principle achieve larger values \cite{Buchler07}.

\section{Vortex-vortex interactions and the fusion degrees of freedom}

In this section we show that in the presence of vortices the spectrum exhibits modes whose energy oscillates and converges to zero with the vortex separation. These modes are argued to correspond to the fusion degrees of freedom of the non-Abelian anyonic vortices. We find the microscopic dependence of the oscillations and show that they are consistent with Majorana fermions being localized at the vortex cores. Finally, we outline the topological low energy spectrum.

\subsection{Oscillating vortex-vortex interactions}

\begin{figure}[t]
\begin{center}
\begin{tabular}{cc} \includegraphics*[width=7cm]{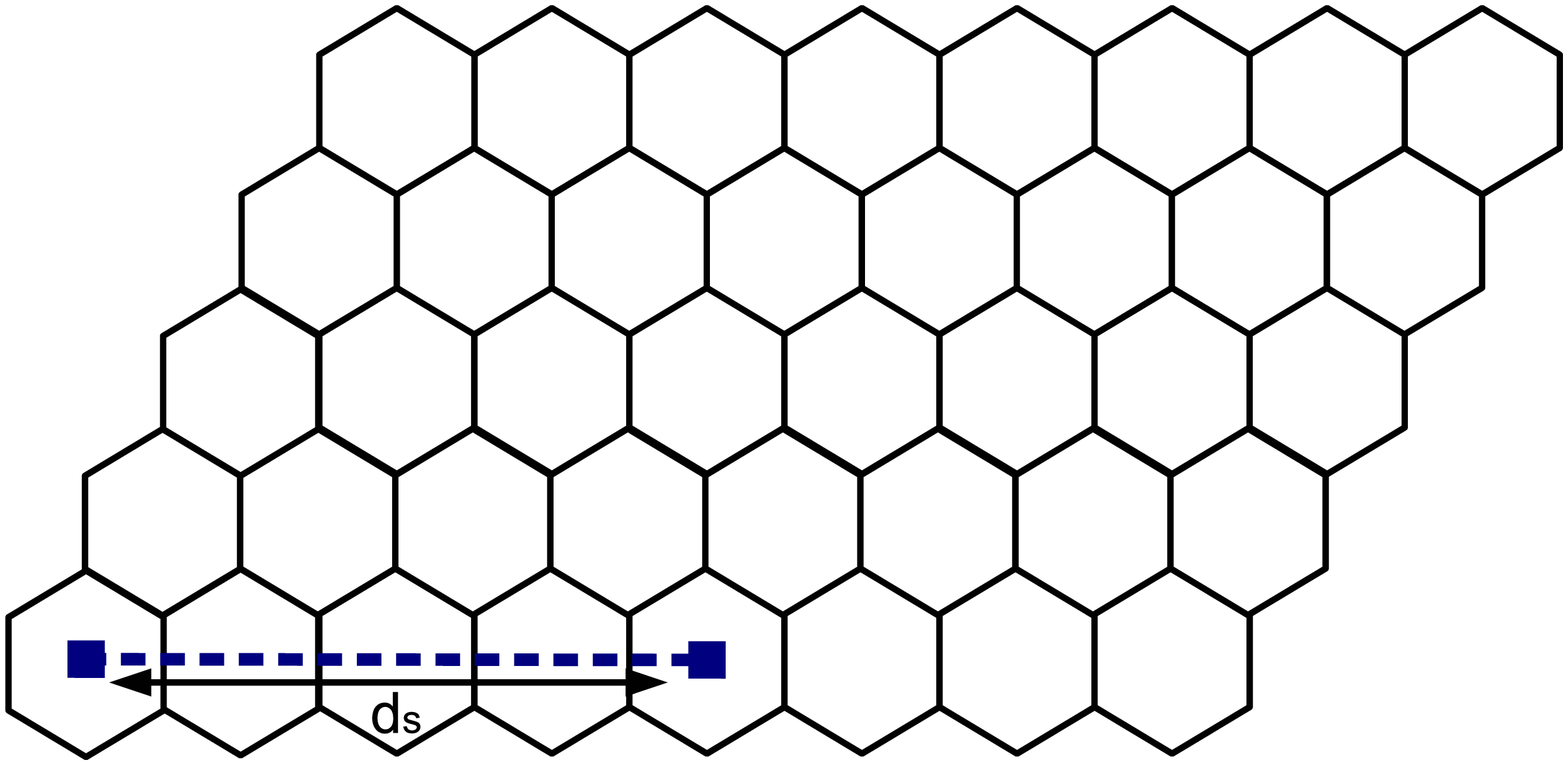} & \includegraphics*[width=7cm]{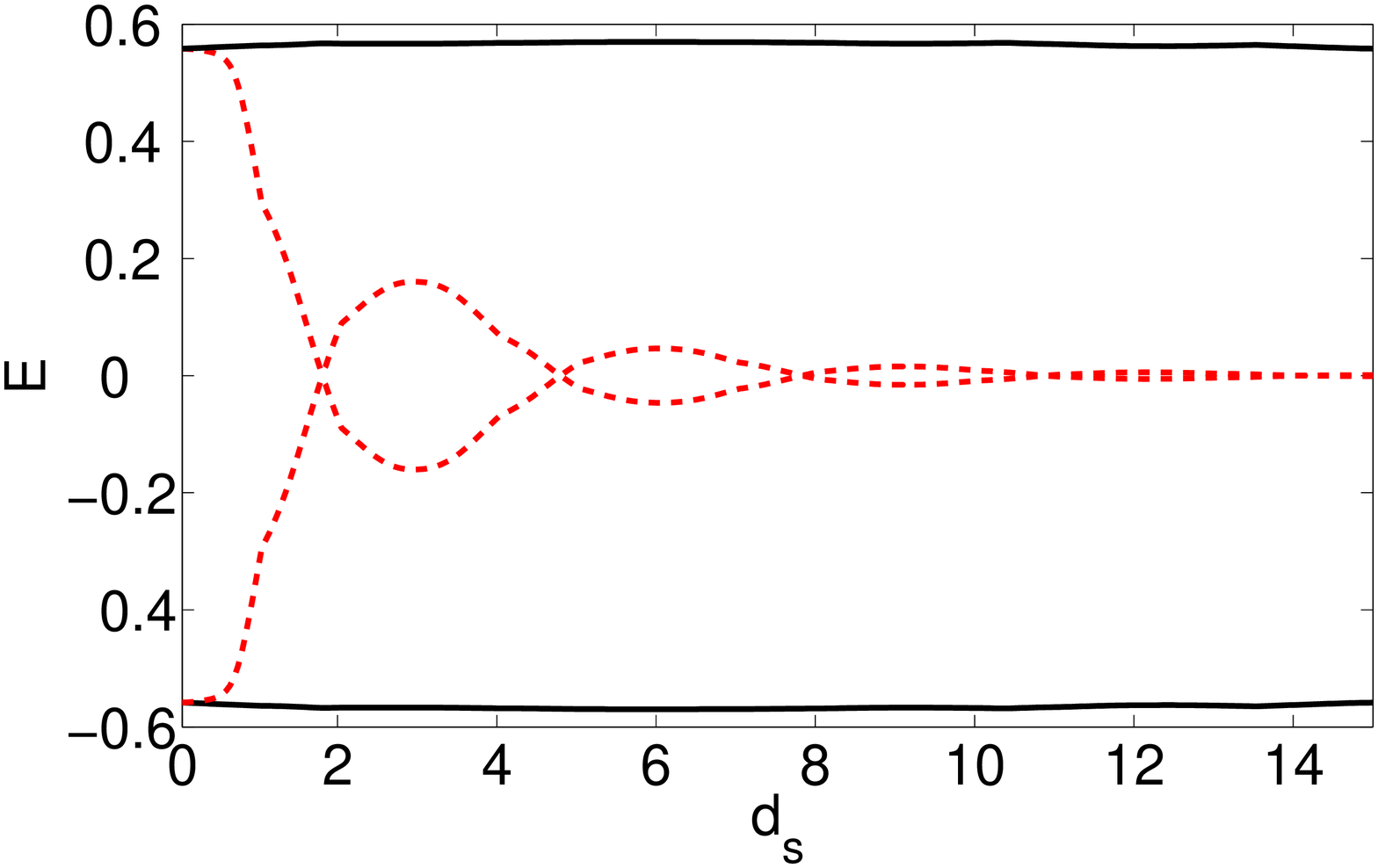} \\ (a) & (b)\end{tabular}
\end{center}
\caption{(a) Two vortex configuration parametrized by the vortex separation $d_s$ (picture not on scale for the $L_x=30$, $L_y=20$ system). The blue squares denote the location of the vortices, which are created by setting $J_{ij}=-1$ on the links crossed by the dashed blue line while $J_{ij}=1$ on all other links. (b) The particle-hole symmetric mode spectrum \rf{Hw} as a function of $d_s$. The two dashed red lines are the oscillating fusion modes with energies $\pm \mathcal{E}^{d_s}$, while the solid black lines are the vortex separation independent free fermion modes with energy gap $\Delta_f$. The plot is for $K=0.05$. }
\label{lattice_2v}
\end{figure}

When two vortices are separated linearly as illustrated in Figure \ref{lattice_2v}(a), the spanned configurations are conveniently parametrized by a continuous vortex separation defined as $d_s = d + \frac{s}{S}$. Here $d$ denotes the number of links of equal coupling strength between the vortices and $1 \leq s \leq S$ is the step when the instantaneous coupling strength on the $(d+1)$th link is $1-\frac{2s}{S}$. At integer values of $d_s$ the vortices are pinned exactly to the plaquettes, while for non-integer values, as illustrated in Figure \ref{transport}(a)-(d), they are interpreted being located at some intermediate position. As $d_s \to 0$, the vortices are brought to the same plaquette. We refer to this as \emph{fusing} the vortices.

\begin{figure}[t]
\begin{center}
\begin{tabular}{cc}
\includegraphics*[width=7cm]{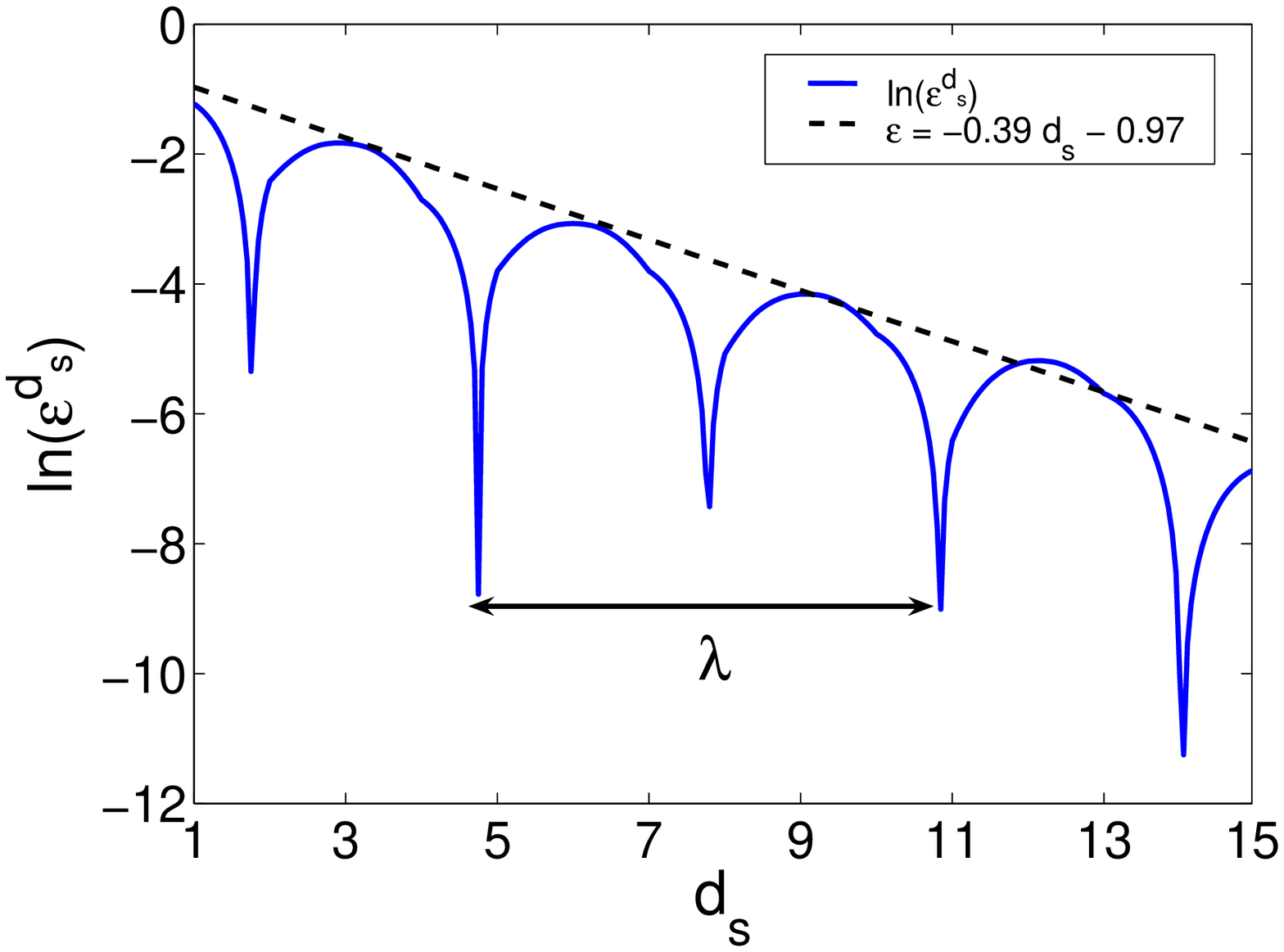} & \includegraphics*[width=6.8cm]{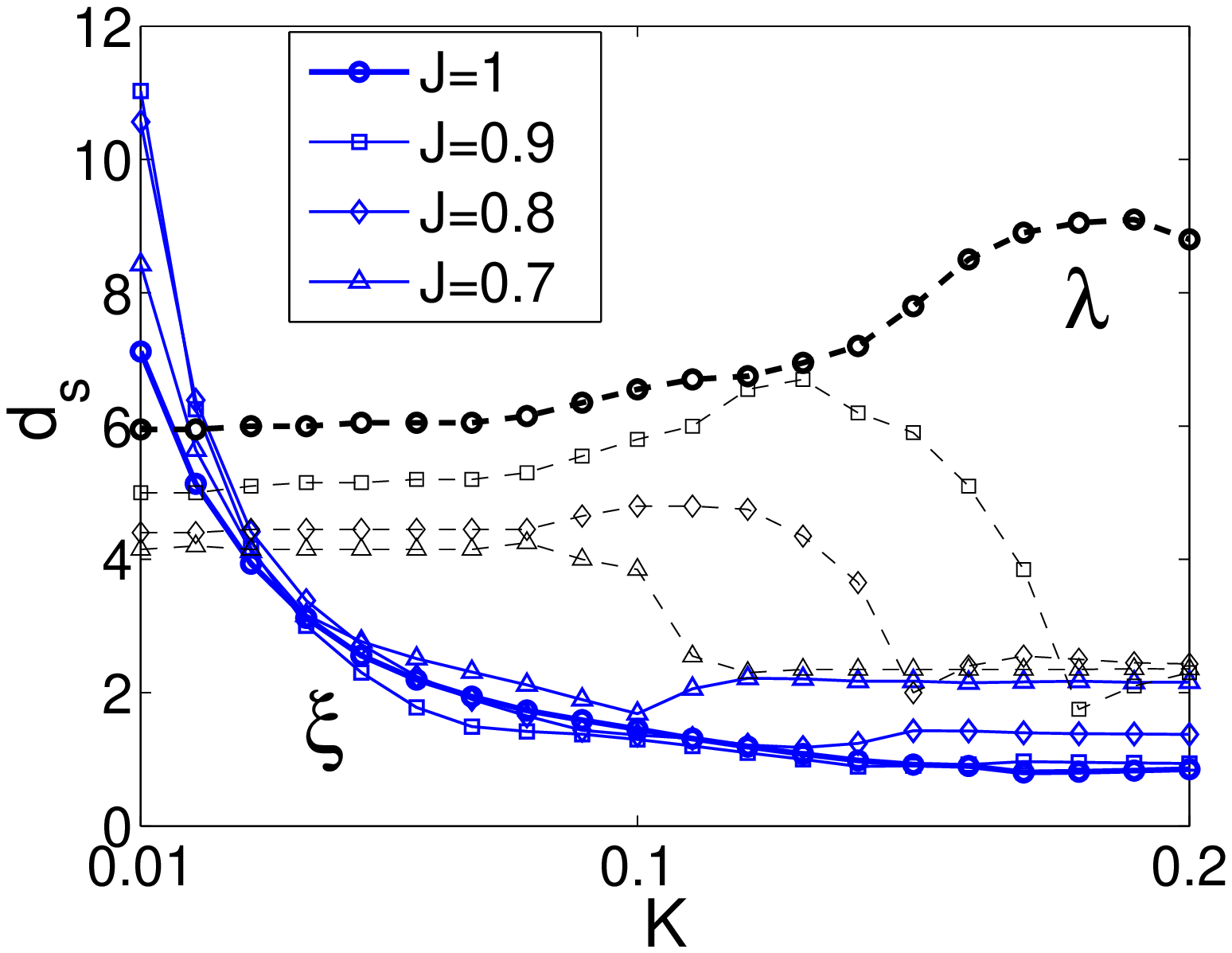} \\
 (a) & (b)
\end{tabular}
\end{center}
\caption{\label{int_2v} (a) A plot of  $\ln (\mathcal{E}^{d_s})$ for $J=1$ and $K=0.05$. A linear fit on the peaks gives $\xi \approx 2.6$ (in units of $d_s$) for the coherence length. The distance between the shown minima gives for the oscillation wavelength $\lambda \approx 6.0$. (b) The coherence length $\xi$ (blue solid line) and the oscillation wavelength $\lambda$ (black dashed line) as functions of $K$ for several $J=J_x/J_z=J_y/J_z$. For $K \lesssim 0.1$ $\xi$ scales as $K^{-1}$, while $\lambda$ remains a $J$ dependent constant. For $K \gtrsim 0.1$ the coherence length $\xi$ converges to a constant, while $\lambda$ acquires vortex separation dependence (this appears as an increase due to $\lambda$ being calculated from fixed nodes at small $d_s$ for which the wavelength first increases), before also converging to a $J$ independent constant.}
\end{figure}

Figure \ref{lattice_2v}(b) shows the spectral evolution as a function of the vortex separation $d_s$. As observed in \cite{Lahtinen08, Lee07}, in the presence of a pair of vortices the energy of the lowest lying mode, $\mathcal{E}^{d_s} \equiv E_1^{d_s}$, converges to zero with vortex separation. The finite energy at small separations is interpreted as being due to short-range vortex-vortex interactions. Higher energy modes are insensitive to it and remain at constant energy $\Delta_f = E_2^{d_s}$ that corresponds to the spectral gap. The continuous nature of the transport, however, reveals the oscillations in the energy $\mathcal{E}^{d_s}$ that have been qualitatively predicted in the context of $p$-wave superconductors \cite{Cheng09}.  Due to particle-hole symmetry, the oscillating modes come in $\pm \mathcal{E}^{d_s}$ pairs, but the crossings are genuine. In general, the oscillating energy can be written as  
\be \label{int}
	\mathcal{E}^{d_s} =  \Delta_f \cos(\frac{2\pi d_s}{\lambda}) e^{-\frac{d_s}{\xi}},
\ee
where non-universal constants $\lambda(J,K)>0$ and $\xi(J,K)>0$ depend on the couplings and parametrize the frequency of the oscillations and the convergence of the energy, respectively. We are especially interested in the latter as it gives the characteristic \emph{coherence length} for particular values of the couplings. Only when $d_s \gg \xi$, one expects the low energy physics to be fully described by the topological Ising anyon theory. The parameters $\lambda$ and $\xi$ can be extracted from the plot of $\ln(\mathcal{E}^{d_s})$ as shown in Figure \ref{int_2v}(a). The linear fit with negative slope confirms the exponential convergence of the energy with vortex separation, and distance between successive dips gives the half of the wavelength. By performing similar linear fits for a range of $J$ and $K$, we obtain Figure \ref{int_2v}(b) that shows $\xi$ and $\lambda$ as functions of the physical parameters. They show very different behavior in the small and large $K$ limits, which roughly occur when $K \lesssim 0.1$ or $K \gtrsim 0.1$, respectively.

Let us consider first the $K \lesssim 0.1$ regime, that is experimentally the more relevant one. There the behavior of both the coherence length and the oscillation wavelength is simple. We find $\xi$ decreasing smoothly as $K^{-1}$, with the $J$ dependence becoming significant only for $K \lesssim 0.01$. This behavior can directly be attributed to the $\Delta \sim K$ scaling of the spectral gap, which implies that the coherence length scales in general as $\xi \sim \Delta^{-1}$. For instance, when $J=1$ the gap is given by $\Delta=6\sqrt{3}K$ (see Figure \ref{lespectrum}(b)), which provides an excellent approximation of the $K \lesssim 0.1$ region in Figure \ref{int_2v}(b). While the coherence length is predominantly controlled by $K$, the oscillation wavelength $\lambda$ depends only on $J$ in this regime. This follows from $\lambda$ being proportional to the inverse Fermi momentum \cite{Cheng09}. To be precise, here it is proportional to the difference of the two Fermi point momenta, i.e. $\frac{2\pi}{\lambda} \sim |\mathbf{p}_F^+-\mathbf{p}_F^-|$. For small $K$ the Fermi momenta are insensitive to the three spin term, while they move away from each other, i.e. $|\mathbf{p}_F^+-\mathbf{p}_F^-|$ increases, when one approaches the phase boundaries ($J \to \frac{1}{2}$) \cite{Lahtinen10}. Thus the observed decrease in $\lambda$ with decreasing $J$ is consistent with this picture.

The behavior in the $K \gtrsim 0.1$ regime is dominated by the behavior of the spectral gap. It converges to a $J$ dependent constant value somewhere in the range $0.1 \lesssim K \lesssim 0.2$ and can not be increased further by increasing $K$ \cite{Lahtinen08}. When this occurs $\xi$ becomes also a constant on the order of the lattice spacing. On the other hand, the oscillation wavelength $\lambda$ enters an intermediate regime where it becomes first dependent on the vortex separation $d_s$. In Figure \ref{int_2v}(b), where the wavelength is calculated from the small $d_s$ behavior, this corresponds to the regime where $\lambda$ first increases and then decreases. When $K$ is increased further, the large $d_s$ behavior takes over and $\lambda$ converges to the value of two lattice constants. This means that for sufficiently large $K$ the oscillations are essentially absent when vortices are pinned to plaquettes (integer values of $d_s$). This behavior is in agreement with the results for $p$-wave superconductors, where changes in the oscillations are known to occur both in the vicinity of phase transitions and for suffiently large gaps \cite{Mizushima10, Cheng10}. Even though the gap saturates in the honeycomb lattice model, the topological phase remains robust even for $K \sim J$ \cite{Jiang11}. We interpret the changes in $\xi$ and $\lambda$ being due to the specific microscopics of the honeycomb lattice model that can modify the non-universal properties of the topological phase. While both the small and large $K$ regimes support the same non-Abelian anyons, they can therefore exhibit very different microscopic signatures depending on the physical parameters. Understanding the microscopics that control these signatures is relevant, for instance, to anyon-anyon interaction driven phase transitions \cite{Ludwig11}, as well as to any topological quantum computing schemes where read-out of information is based on detecting the energy shifts of the topological states \cite{Cheng09}.

\subsubsection{Rotational anisotropy and transversal transport}

\begin{figure}[t]
\begin{center}
\begin{tabular}{ccc} 
\includegraphics*[width=4.5cm]{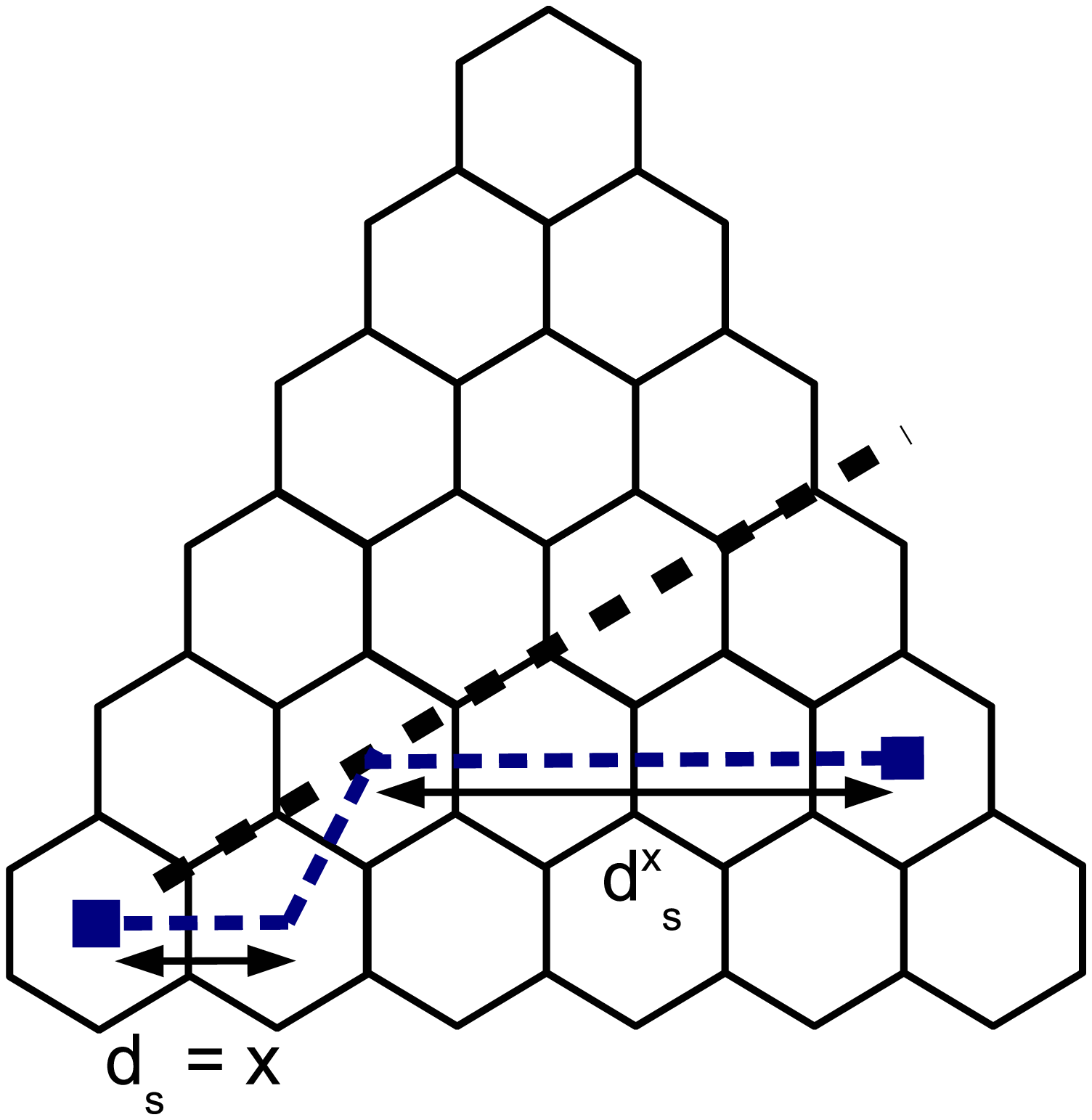} & \includegraphics*[width=4.5cm]{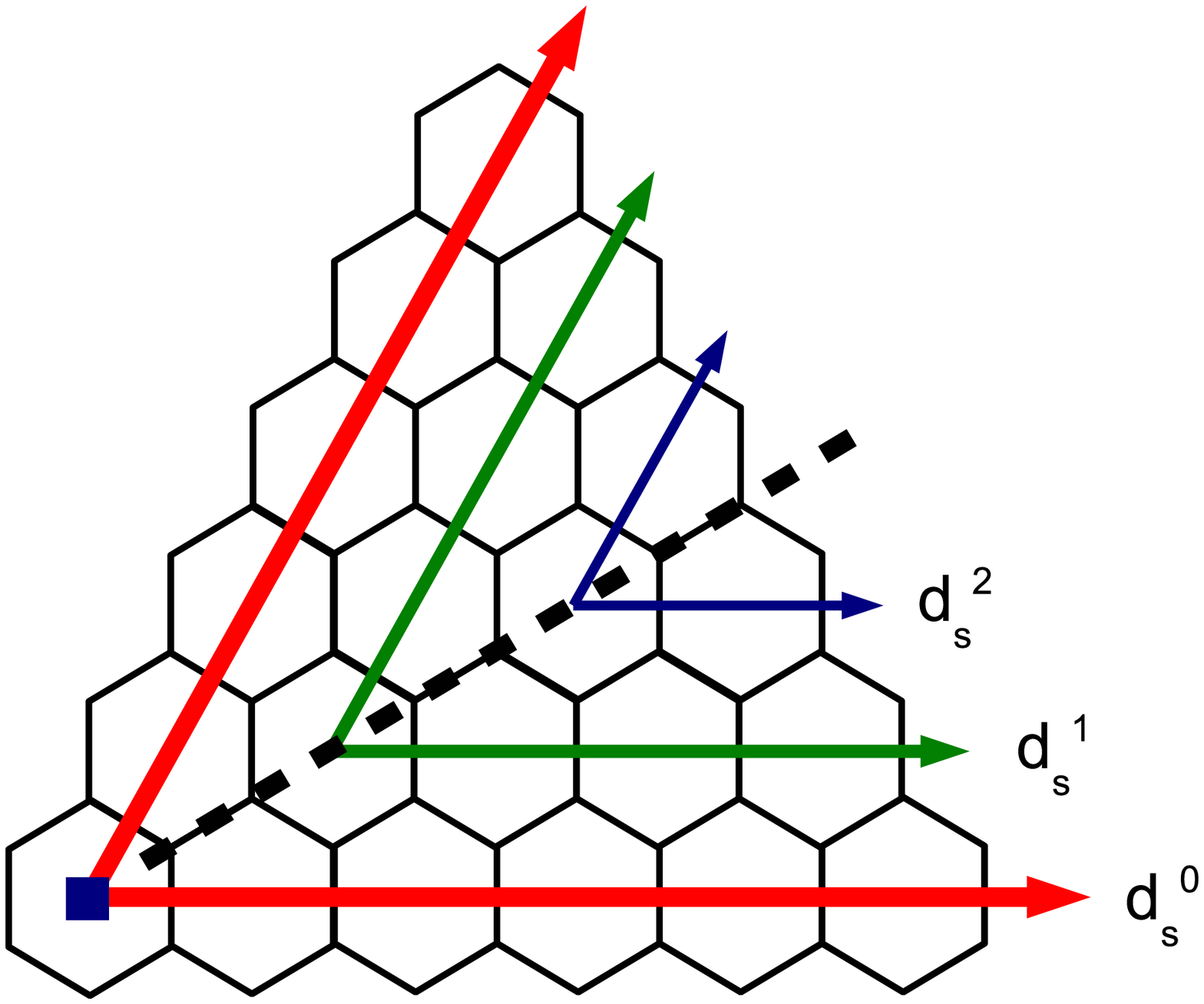} & \includegraphics*[width=5.5cm]{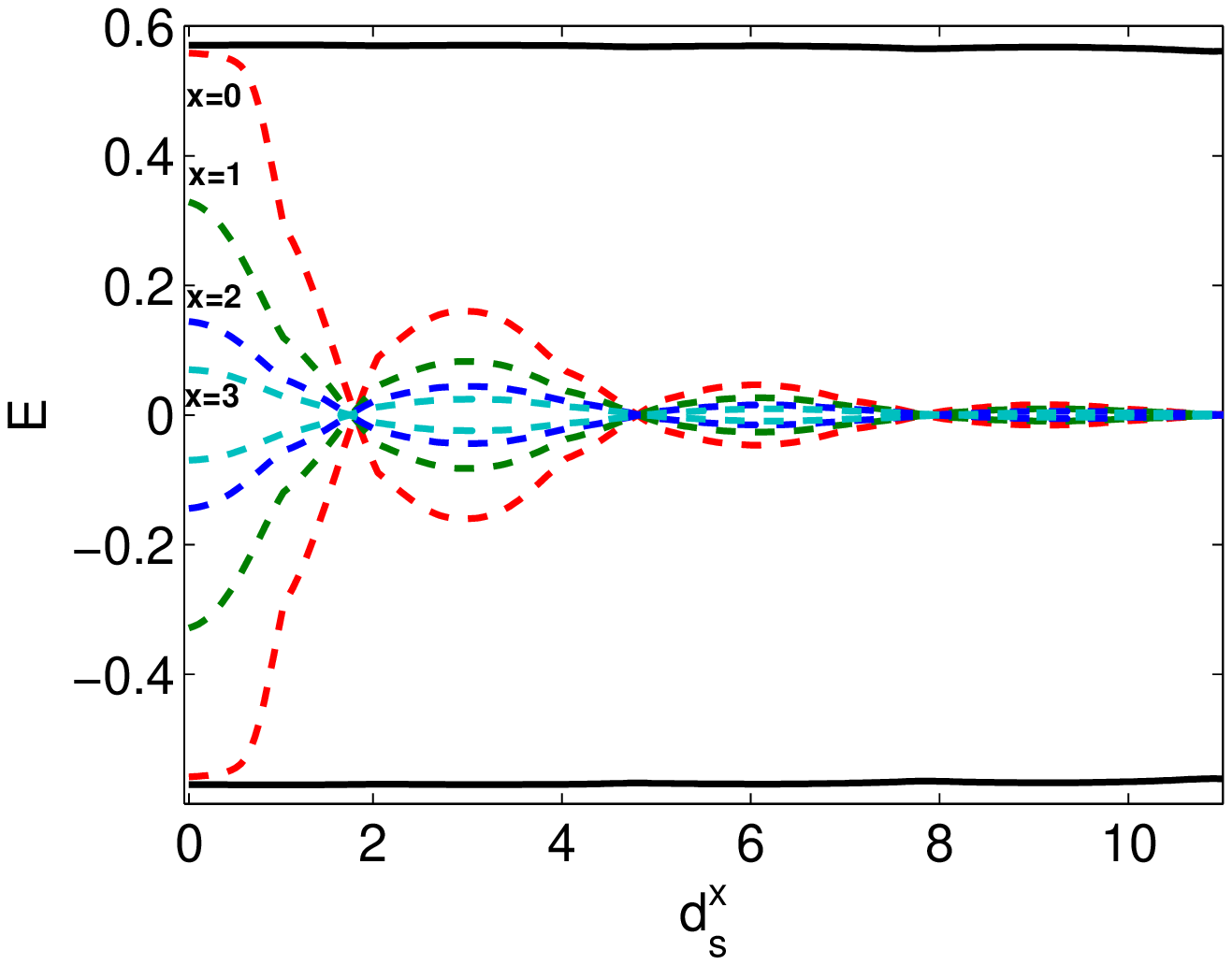} \\
(a) & (b) & (c)
 \end{tabular}
\end{center}
\caption{ (a) Transportation of a vortex at separation $x$ away from the sector boundary (black dashed line). (b) When a vortex is fixed at the origin located at the lower left corner and the other is transported along the arrows, the fusion mode energy behaves identically along the arrows of same colour. Parallel arrows form altogether six sectors with identical behavior. The picture shows parts of two sectors that are separated by a boundary in direction $\mathbf{n}_x + \mathbf{n}_y$. (c) When vortices are moved in parallel directions as shown in (b), the energy oscillates with the same wavelength. The plot colours correspond to the arrows of same colour in (a). }
\label{sectors}
\end{figure}

To characterize the fusion mode energy $\mathcal{E}^{d_s}$ for arbitrary relative vortex locations, we consider two further transport processes. First, let us consider the spectral evolution for the transport shown in Figure \ref{sectors}(a) for various orientations of the lattice. If we assume the vortex at the lower left corner to be located at the origin, we find energy behaving identically in six sectors bounded by the directions $\pm (\mathbf{n}_x + \mathbf{n}_y)$, $\pm (\mathbf{n}_x - 2\mathbf{n}_y)$ and $\pm (2\mathbf{n}_x - 2\mathbf{n}_y)$. Figures \ref{sectors}(b) and \ref{sectors}(c) show that within these sectors one can define parallel directions along which the oscillations have always the same wavelength with the nodes coinciding. The only difference is the overall exponential damping with the vortex separation when the transport is carried out further away from the origin. This sector structure contrasts with the rotational isotropy in $p$-wave superconductors \cite{Gurarie07}. The difference is again in the microscopics, which this time derives from to the underlying honeycomb lattice.

\begin{figure}[t]
\begin{center}
\begin{tabular}{cc}
\includegraphics*[width=6cm]{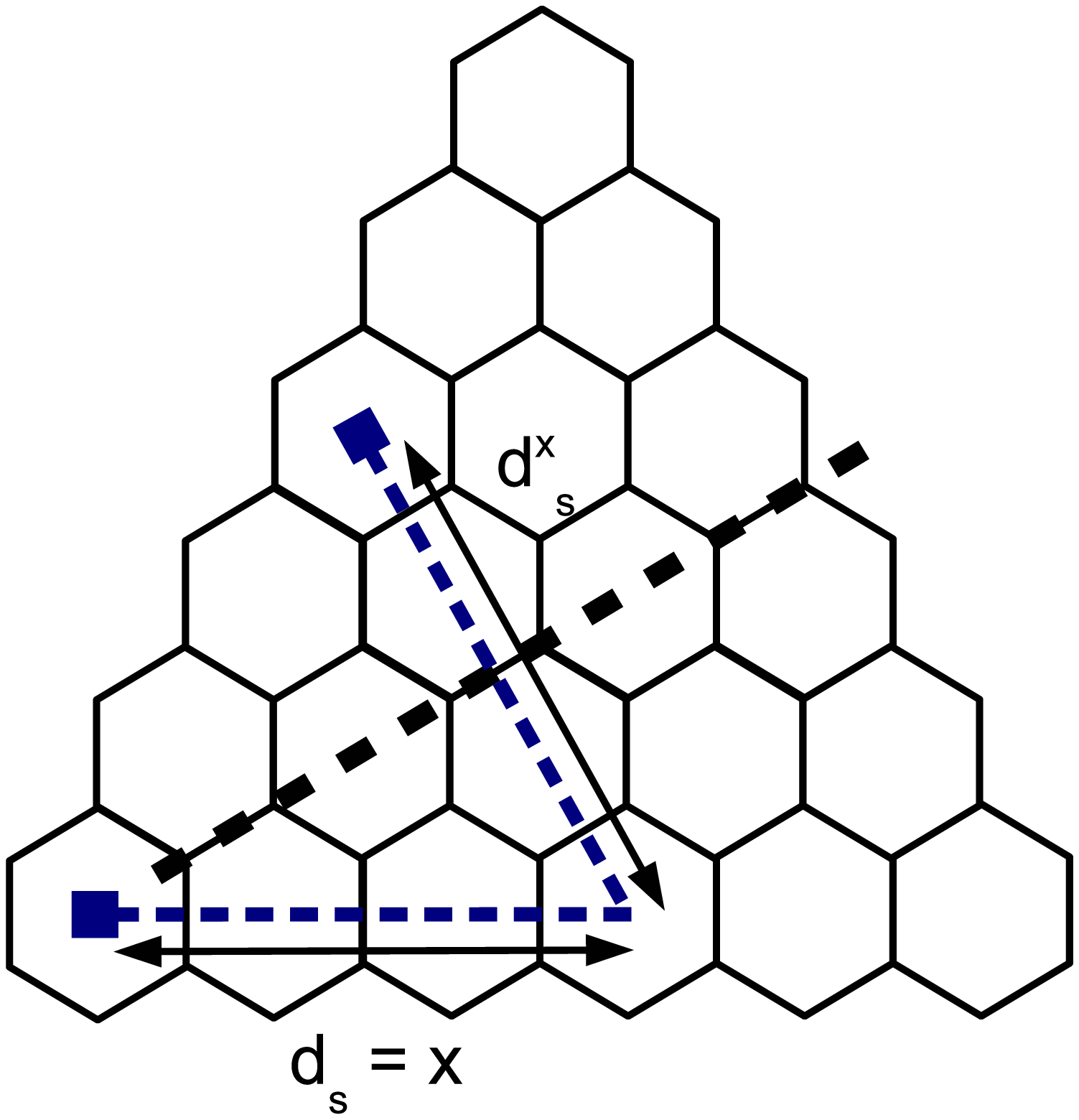}  & \includegraphics*[width=7cm]{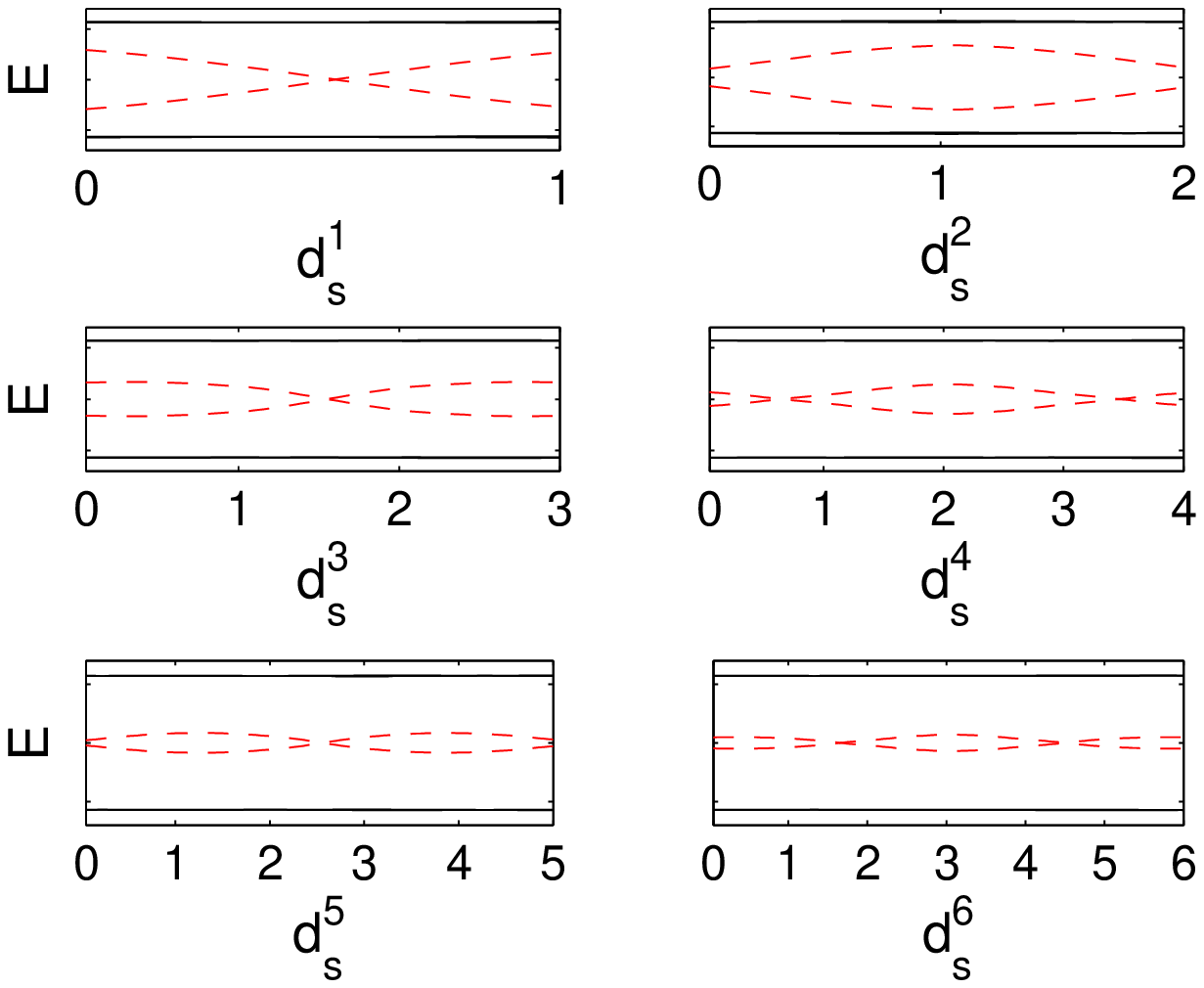} \\
(a) & (b) 
\end{tabular}
\end{center}
\caption{(a) A parametrization for transversal transport. Starting with a configuration of two vortices separated linearly by $x$ plaquettes, we define the transversal direction to be $-\mathbf{n}_x+\mathbf{n}_y$. $d_s^x$ denotes the distance from the initial configuration. (b) For an even $x$ the energy splitting may oscillate, but one always arrives at the same sign of the splitting. For an odd $x$ the sign always changes such that the energy splitting is exactly zero at the sector boundary.}
\label{lattice_diag}
\end{figure}

Finally, we consider the fusion mode energy when vortices are transported transversally with respect to each other as illustrated in Figure \ref{lattice_diag}(a). As the oscillations are known to behave identically along the sector centers, one expects the energy to be reflection symmetric with respect to the boundaries. This is indeed the case as shown in Figure \ref{lattice_diag}(b) for various distances $x$ from the origin, but we also find that the sign of the energy may change. For an even $x$ the energy may oscillate, but one always arrives at the same sign, while for an odd $x$ the sign always changes such that the energy is exactly zero at the sector boundary. This implies that one can arrive at different conclusions on the energy depending on the path chosen to transport the vortices. As all physical vortex states have to be gauge invariant, i.e. not depend on the path \cite{Kitaev05}, this sets limits on simulating vortex transport through manipulating the couplings. Only tuning the couplings along paths without an ambiguity simulate the evolution of the physical states. This can be achieved by considering always the shortest path, which corresponds to tuning only the $J_y$ and $J_z$ couplings. The restriction to two out of the three link types arises also naturally in the fermionization techniques that do not require additional gauge symmetrization \cite{Kells09, Chen08}.

\subsection{The low energy spectrum and the fusion degrees of freedom}

\begin{figure}[t]
\begin{center}
\begin{tabular}{cc}
 \includegraphics*[width=7.7cm]{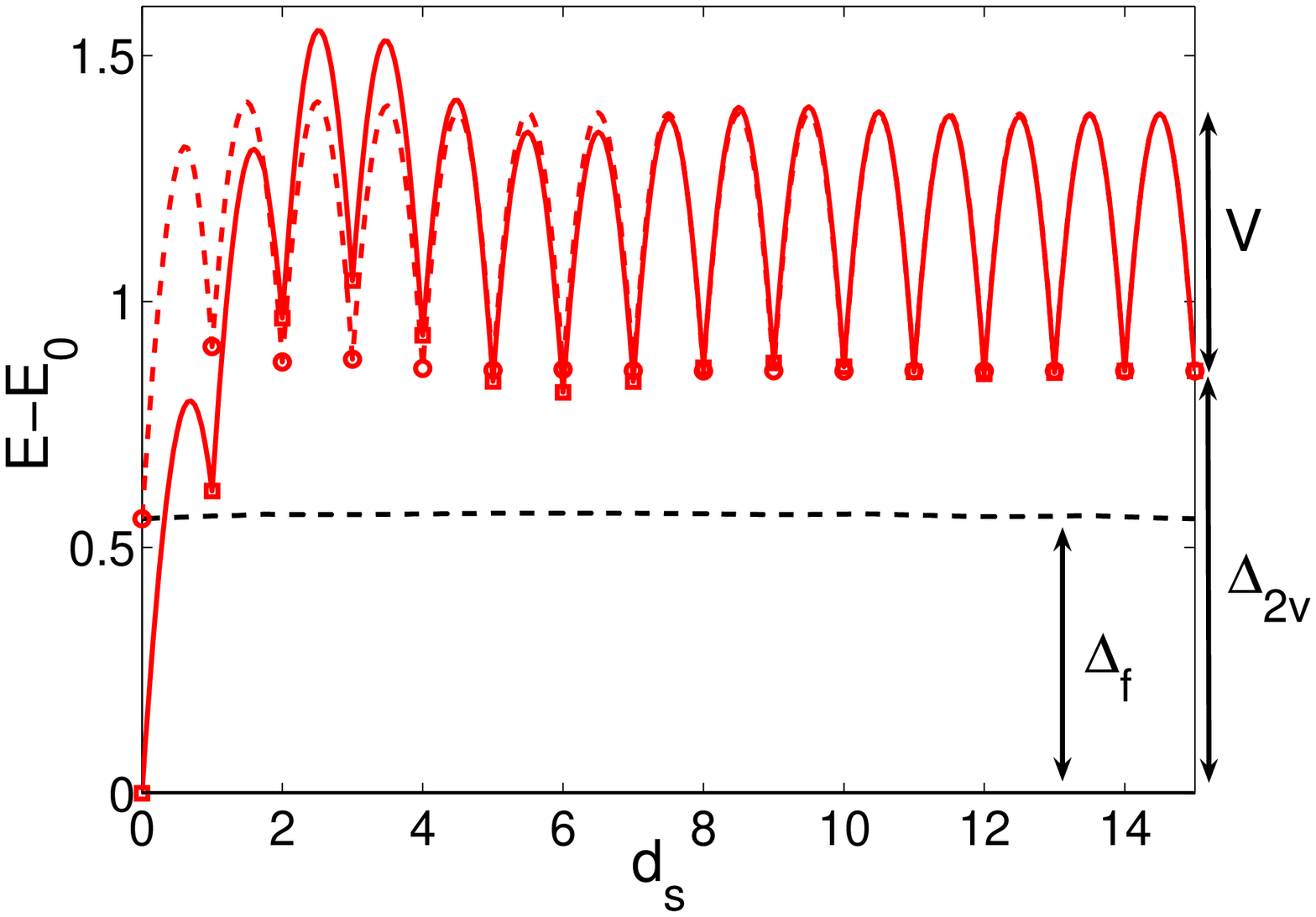} & \includegraphics*[width=7.2cm]{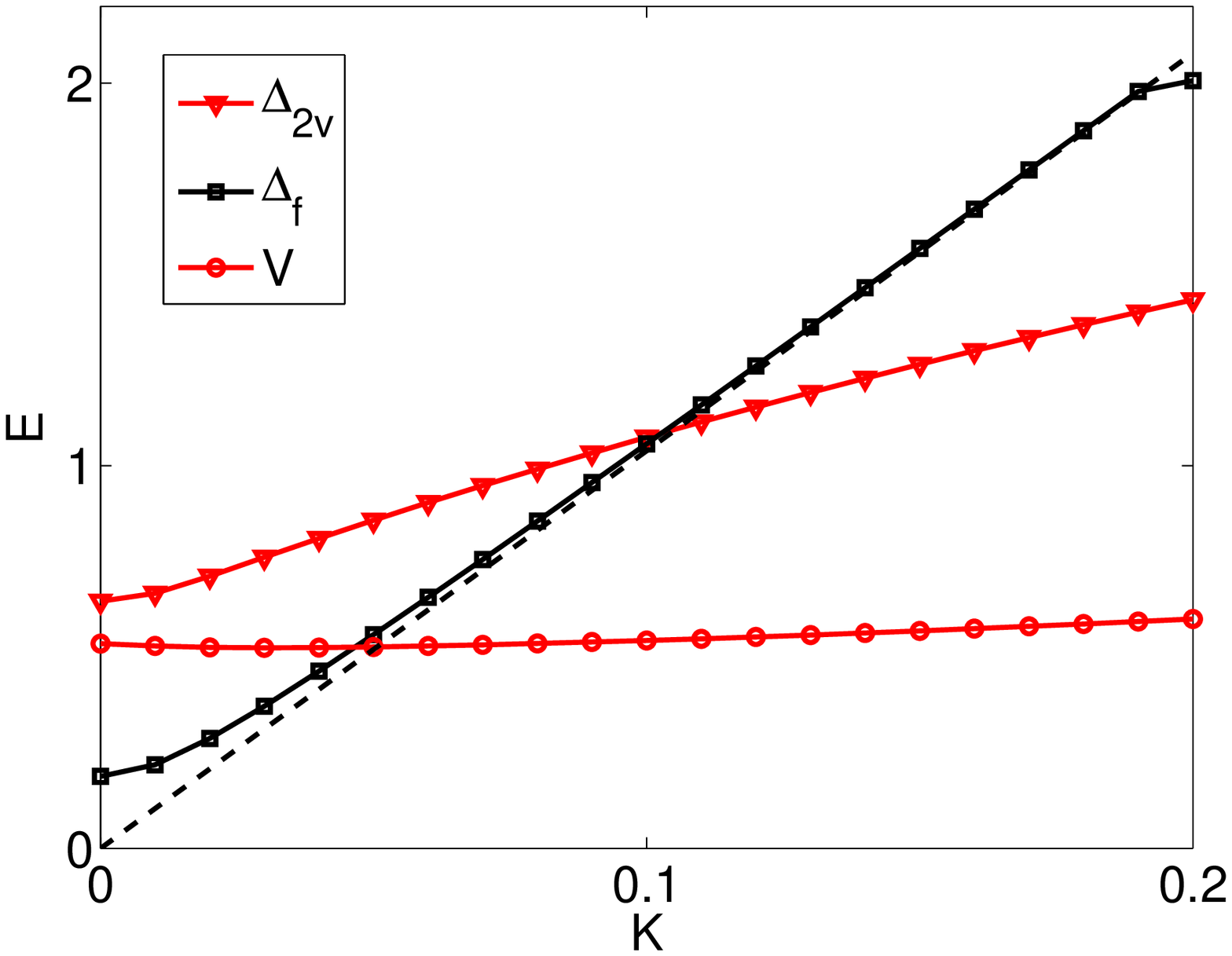} \\
 (a) & (b) 
\end{tabular} 
\end{center}
\caption{(a) The low-energy spectrum of the two vortex sector as a function of $d_s$ for $J=1$ and $K=0.05$. All energies are with respect to $E_0$, the ground state energy of the vortex-free sector. The dashed black line is the fermion gap $\Delta_f$ in the absence of vortices. The solid (squares) and dashed (circles) red lines are the exponentially degenerate two vortex sectors states with energies $E^{d_s}_0$ and $E^{d_s}_0+\mathcal{E}^{d_s}$, respectively, that correspond to an unoccupied and an occupied fusion mode. The markers denote positions	 when the vortices are located exactly at plaquettes. (b) The fermion gap $\Delta_f$, the vortex gap $\Delta_{2v}$ and the lattice potential $V$ as functions of $K$. The dashed line is the analytic solution for the fermion gap \cite{Kitaev05}, that in the presence of vortices vanishes at $K=0$ only in the thermodynamic limit. For $K>0.2$ the fermion gap becomes a constant $\Delta_f \approx 2.0$. }
\label{lespectrum}
\end{figure}

It has been argued in \cite{Lahtinen08} that the modes having the energy behavior \rf{int}, to be referred to here as \emph{fusion modes}, should be identified with the anyonic fusion degrees of freedom. We explicitly verify this by plotting in Figure \ref{lespectrum}(a) some of the lowest lying states in the two vortex sector with respect to the lowest lying states in the vortex-free sector ($w_p=1$ on all plaquettes). At large $d_s$ there are two exponentially degenerate states with energies $E_0^{d_s}$ and $E_0^{d_s}+\mathcal{E}^{d_s}$ that differ by the occupation of the fusion mode. Here $E_0^{d_s}$ denotes the energy of the state with unoccupied fusion mode. We define it by
\be
	E_0^{d_s} = - \sum_{i=2}^{L_x L_y} E_i^{d_s}-\mathcal{E}^{d_s}, \label{Egs}
\ee
as the state with the lowest energy when vortices are nearby. It follows from this definition, that when the vortices are brought closer and finally fused, the energy of the state with the unoccupied fusion mode first oscillates and then evolves smoothly to $E_0$, the ground state of the vortex-free sectors. On the other hand, the energy of the state with the the occupied fusion mode becomes $\Delta_f$, the first excited free fermion state on the vortex-free sector. In the language of topological field theories \cite{Rowell09,Kitaev05}, the vortex-free ground state, the vortices and the fermions carry a topological quantum numbers $1$, $\sigma$ and $\psi$, respectively. They satisfy the fusion rule $\sigma \times \sigma = 1 + \psi$, which tells that there are two alternatives how a pair of vortices can behave when they are fused: they can either annihilate to the vacuum or leave behind a fermion. Figure \ref{lespectrum}(a) demonstrates that this fusion degree of freedom is exactly that of the occupation of the fusion modes\footnote{Strictly speaking there is no ground state degeneracy in the two vortex sector due to fermionic parity conservation. However, the same study could have been carried out with another vortex pair in the system, in which case it is possible to define degenerate states of same fermionic parity. As long as these additional vortices were far away from the other, this would not affect the spectrum and the presented analysis would apply identically.}. We also observe that when the vortices are fused, the vacuum channel is energetically favoured. This contrasts with the behavior of quasiholes in the Moore-Read state, for which one finds the $\psi$ fusion channel to be energetically favoured \cite{Baraban09}. These different results highlight the different microscopics of the models, that can dramatically affect the non-universal behavior of the topological phases. 

While the honeycomb lattice model does not accommodate analytic treatment of the fusion modes, their origin can be understood in the context of $p$-wave superconductors to which the honeycomb lattice model can be mapped \cite{Yu08,Kells09,Chen08}. There one can explicitly show that vortices bind unpaired massless Majorana fermions \cite{Read00}, whose wave functions exhibit exponentially damped oscillating tails \cite{Gurarie07}. The width of the wave functions is controlled by the fermion gap $\Delta_f$ that can be viewed as a height of a potential well confining them to the vortex cores \footnote{
This is also in agreement with the general theory that says that the fusion channel degeneracy lifting can always be understood as virtual exchange of a fermionic $\psi$ quasiparticle \cite{Bonderson09}. As $\Delta_f$ gives their minimal energy, such processes become less probable as the fermions become more massive.}. When two vortices are nearby each other, the overlap of these tails results in a finite tunneling amplitude between vortex cores, which in turn lifts the degeneracy and gives rise to an oscillating finite energy \cite{Cheng09}. Our results of the oscillating fusion mode energies are consistent with this picture. While the underlying lattice makes it hard to visualize the oscillations in the wavefunctions, by taking appropriate linear combinations of the fusion modes one can find Majorana modes that have support only on the sites around an isolated vortex \cite{Kells10}. In Section 4 we will employ this picture of localized Majorana modes to construct an effective lattice model for studying degeneracy lifting in many vortex systems.

\subsection{Energy gaps and stability}

When the vortices are well separated, the topological low energy spectrum of Figure \ref{lespectrum}(a) can be characterized by the following energy gaps:
\bq
	\Delta_f & = & E_{2}^{d_s}, \label{fgap} \\
	\Delta_{2v} & = & \lim_{d_S \to \infty} E^{d_S}_0 - E_0, \label{vgap} \\
	V & = & \lim_{d_S \to \infty} E^{d_{S/2}}_0-E^{d_S}_0. \label{V}
\eq
$\Delta_f$ and $\Delta_{2v}$ are the fermion and vortex gaps, respectively, that describe the minimal energy required to excite a free fermion mode and a pair of non-interactiong vortices. $V$ is a local potential that favours the vortices to be located at the plaquettes. To study their physical parameter dependence, we plot them in Figure \rf{lespectrum}(b) as functions of the effective magnetic field $K$. We observe a crossing in the magnitudes of the energy gaps such that for $K<0.1$ ($K>0.1$) there holds $\Delta_f < \Delta_{2v}$ ($\Delta_f > \Delta_{2v}$). This means that while for all values of $K$ the vacuum channel is energetically most favoured, for $K<0.1$ the energy of the system can also be lowered by fusing the vortices into fermions. For large effective magnetic field ($K>0.2$) the energy gap becomes a constant, the exact value depending only on $J$ \cite{Lahtinen08} ($\Delta_f \approx 2.0$ for $J=1$), and can no longer be made larger by increasing $K$.

The periodic ``hopping'' of the energies as a function of $d_s$ has not been observed before. The minima always occur for integer values of $d_s$, i.e. when vortices are pinned to plaquettes, whereas the maxima occur at $d_{S/2}$, i.e. when the transported vortex occupies a composite plaquette twice the size of a regular plaquette. We interpret this energy required to move a vortex to an adjacent plaquette as a local potential of magnitude $V$, defined asymptotically by \rf{V}. As shown in Figure \ref{lespectrum}(b), it varies only slightly with $K$ and can thus be effectively treated as constant for a given uniform coupling configuration $J$. Therefore, the vortices can be viewed as living on a uniform periodic background potential whose minima coincide with the sites of the dual triangular lattice. 

Together the energy gaps $\Delta_f$ and $\Delta_{2v}$ and the potential $V$ give a measure of the stability of the topological low energy theory against local perturbations. When the perturbations are weaker than $\Delta_f$ and $\Delta_{2v}$, the spontaneous creation of both fermion and vortex excitations is suppressed. When they are also weaker than $V$, any existing vortices in the system will remain stationary. $\Delta_f$ describes the stability of the topological phase itself, whereas $\Delta_{2v}$ and $V$ describe stability of the vortex sectors within it.

\section{Degeneracy lifting in many vortex systems} \label{section_eff}

In this section we study degeneracy lifting in many vortex systems. Employing an effective lattice model of Majorana fermions, we show that the hybridized spectrum of the fusion modes can be understood in terms of bi-partite interactions between the vortices.

\subsection{Free Majorana fermions on a lattice}

\begin{figure}[t]
\begin{center}
\begin{tabular}{cc} \includegraphics*[width=6cm]{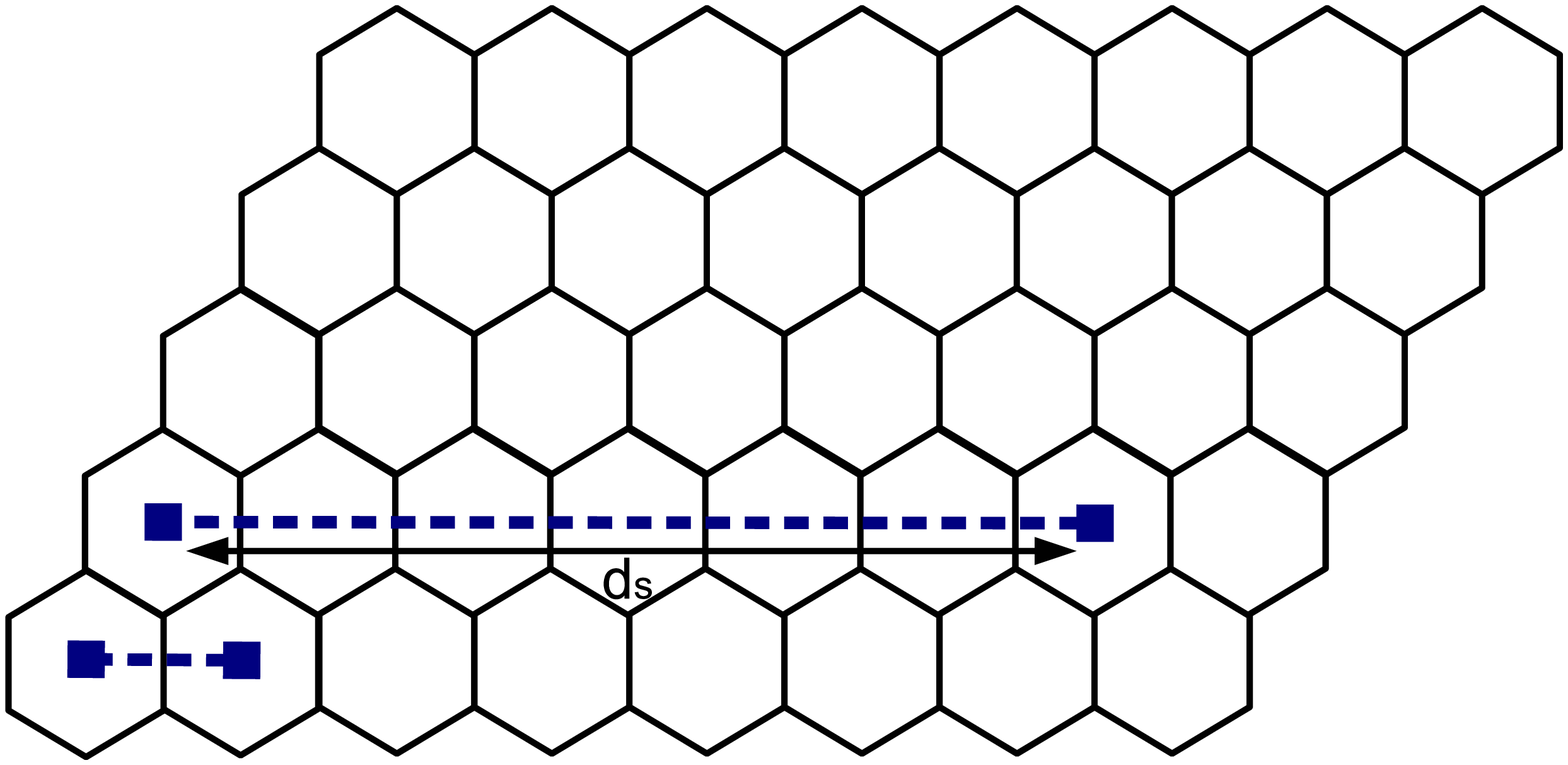} & \includegraphics*[width=6cm]{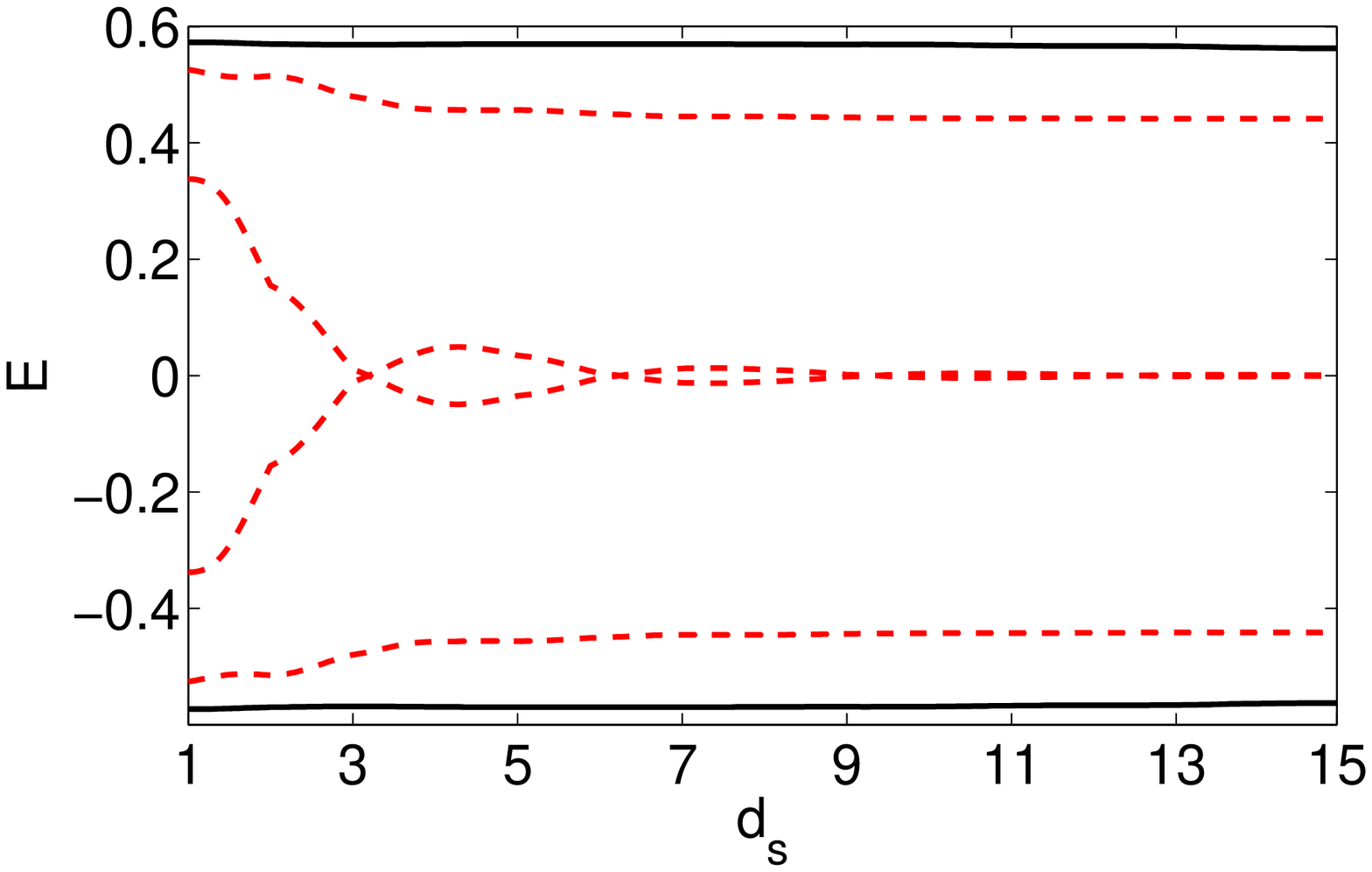} \\
(a) & (b) \end{tabular}
\end{center}
\caption{ (a) A four vortex configuration with one vortex separated by $d_s$ from the other three. (b) One of the two fusion modes oscillates and converges quickly to zero energy, while the other remains localized at finite energy on the group of three vortices. }
\label{lattice_4v}
\end{figure}

In the previous section we analyzed the spectral evolution in two vortex systems. A simple generalization is to consider what happens when a single vortex is dragged away from a bunch of vortices. When this is performed for the four vortex configuration shown in Figure \ref{lattice_4v}(a), one still finds qualitatively similar behavior. Figure \ref{lattice_4v}(b) shows that one of the fusion modes to oscillates and converges to zero energy, while the other remains at finite energy. In other words, one of the two fusion modes is delocalized while the other remains localized at the three vortices. However, comparing to Figure \ref{lattice_2v}(b), the energy of both fusion modes, including the amplitude and the wavelength of the oscillations, is modified due to the presence of additional vortices. As more vortices are added, the fusion modes will in general hybridize a new energy band within the spectral gap \cite{Lahtinen10}.

\begin{figure}[t]
\begin{center}
\begin{tabular}{ccc} 
\includegraphics*[width=4cm]{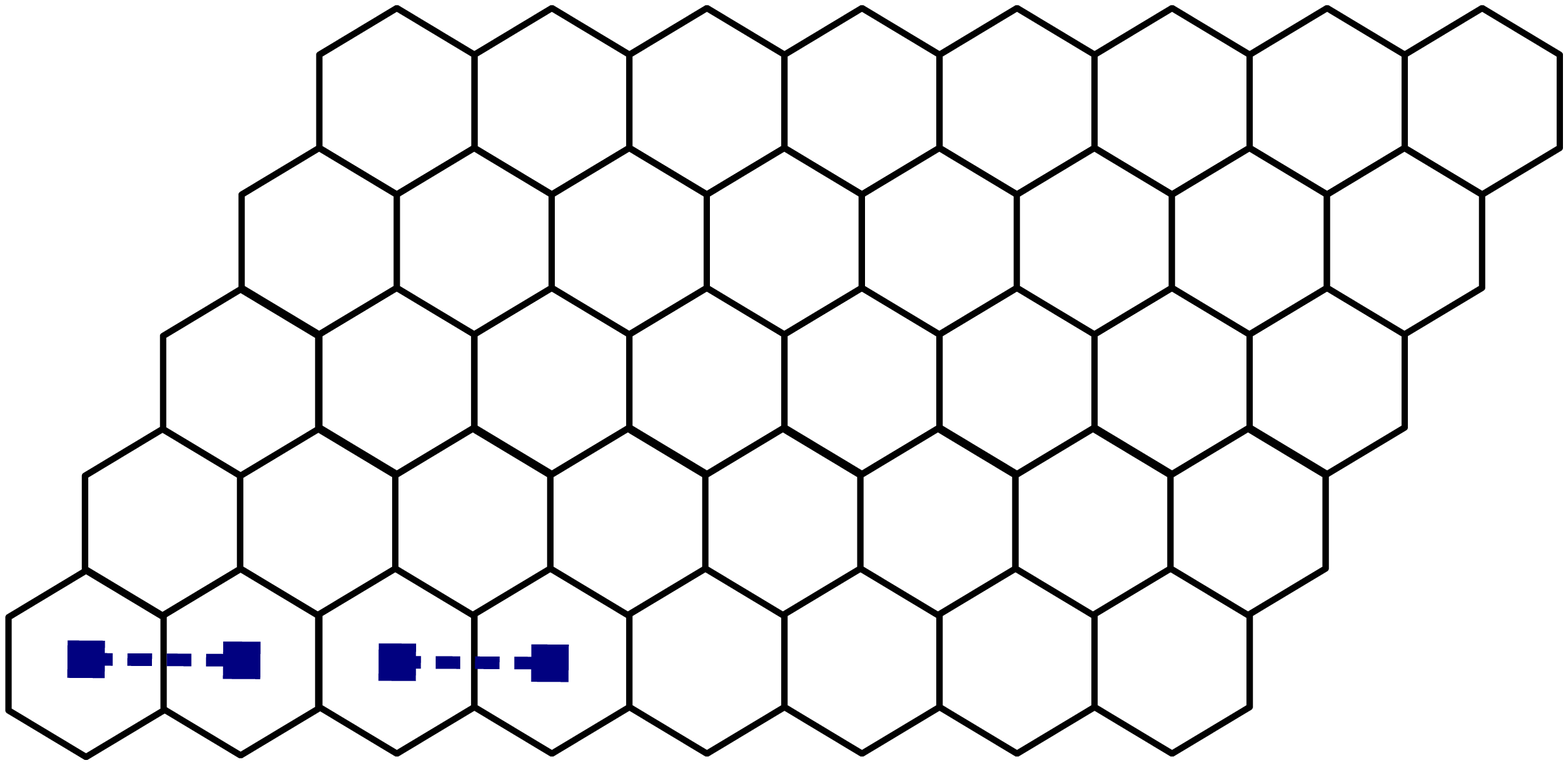} & \includegraphics*[width=4cm]{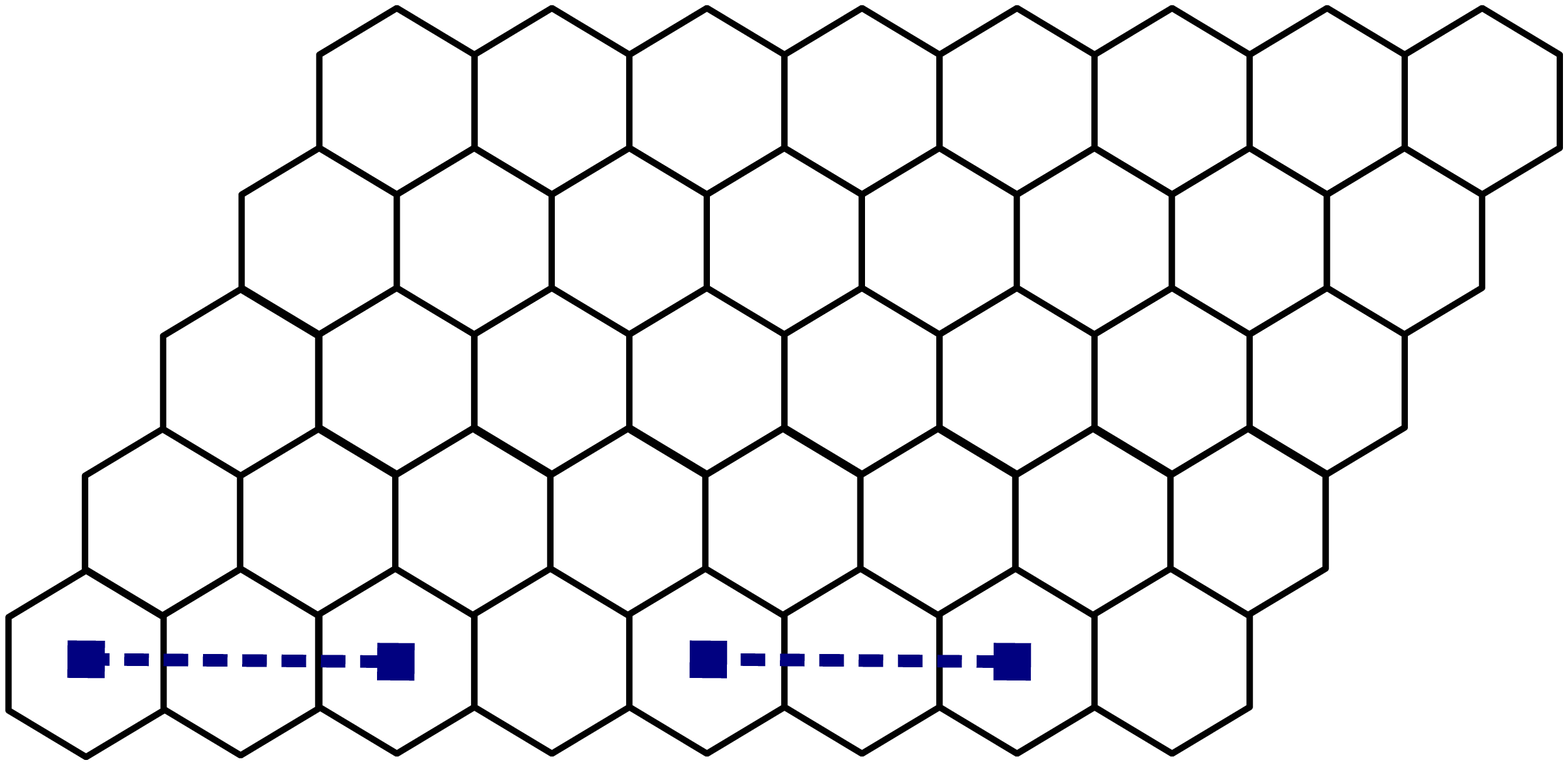} & \includegraphics*[width=4cm]{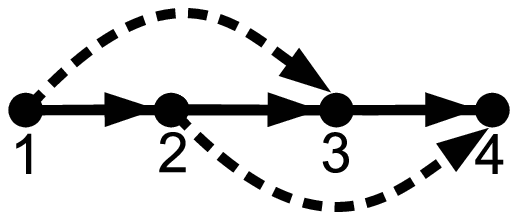} \\
(a) & (b) & (c) \\
\includegraphics*[width=4cm]{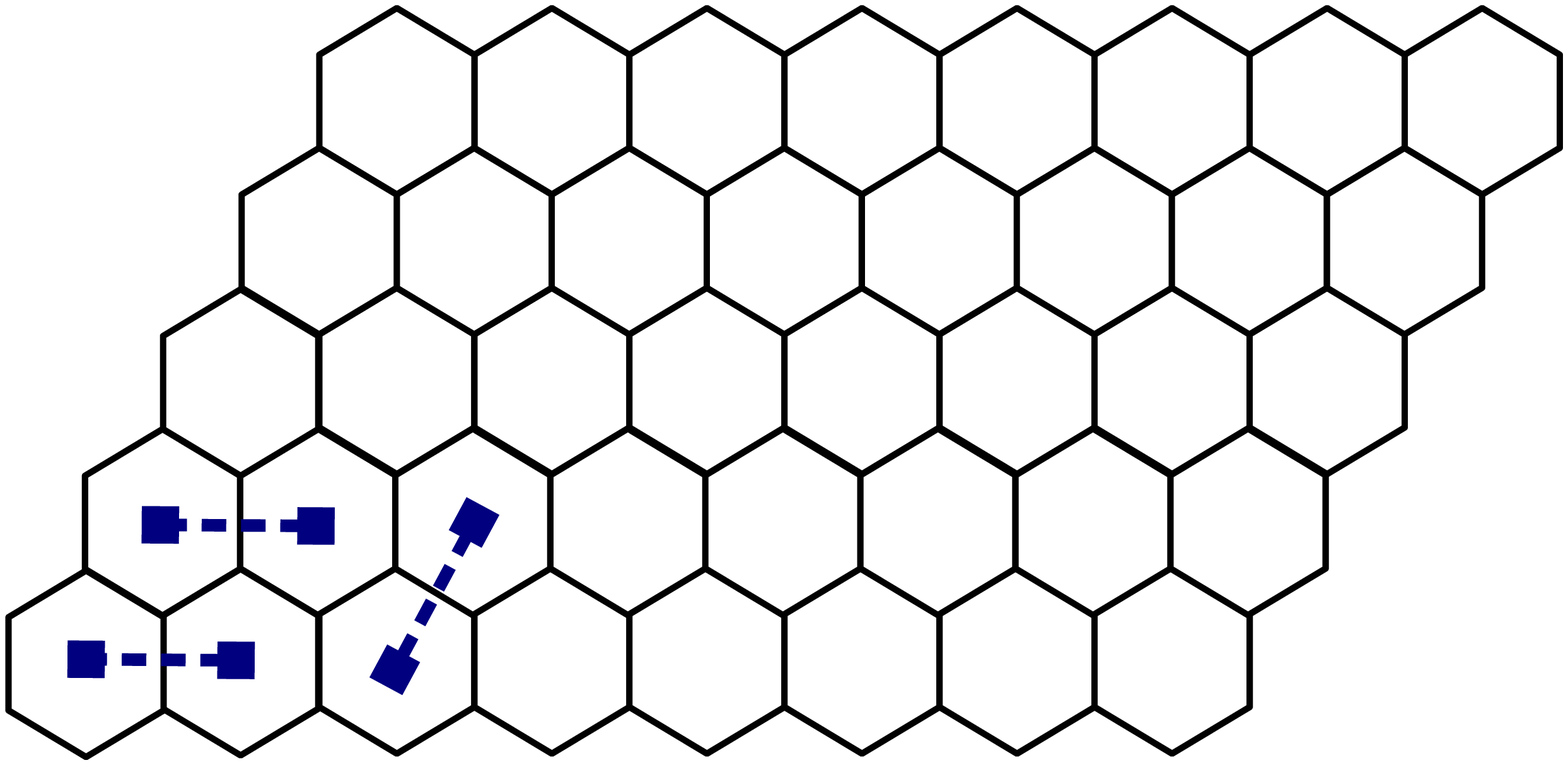} & \includegraphics*[width=4cm]{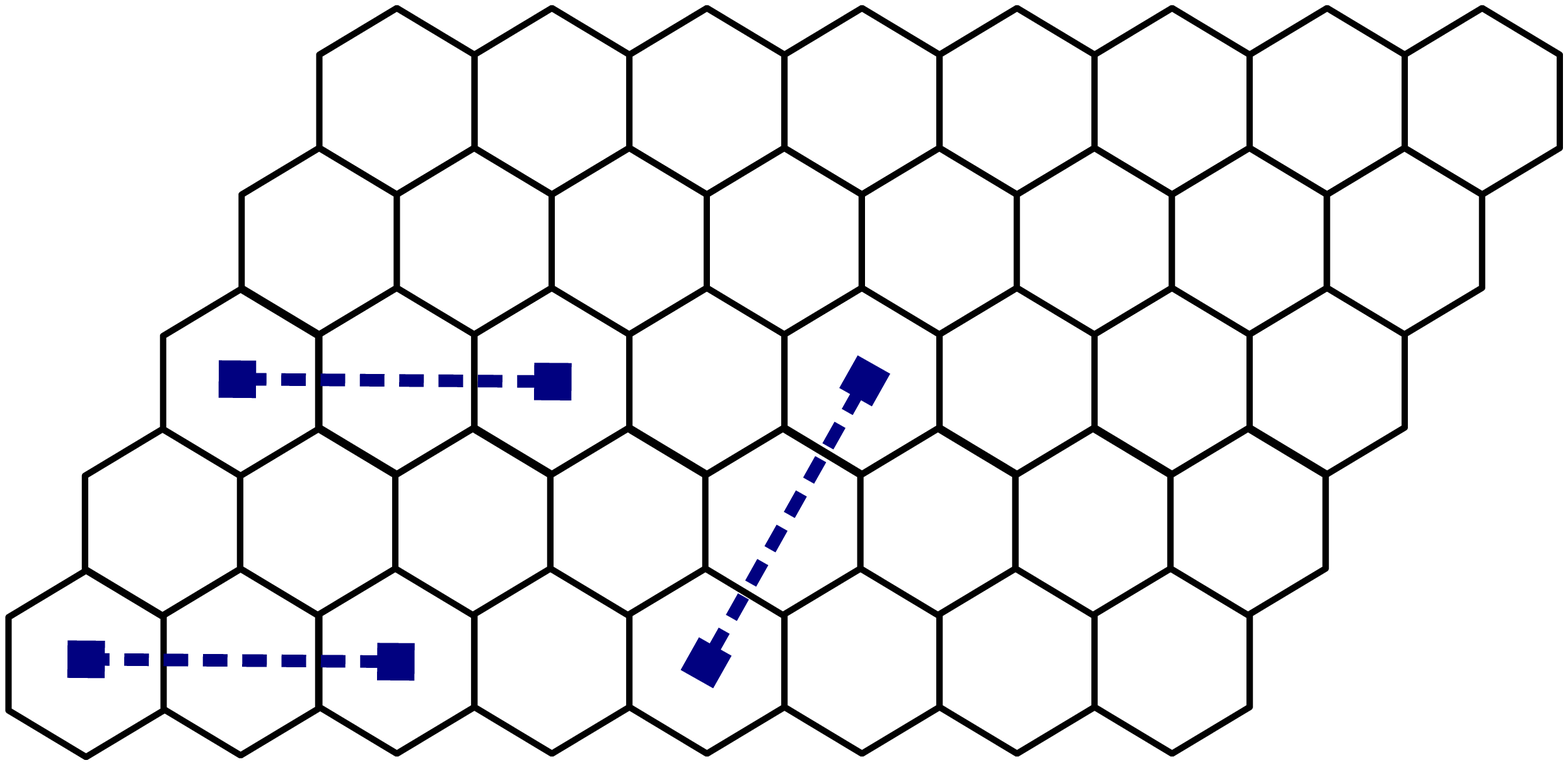} & \includegraphics*[width=4cm]{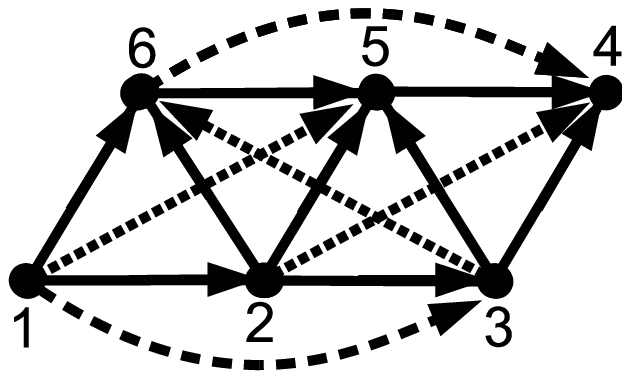} \\
(d) & (e) & (f)
 \end{tabular}
\end{center}
\caption{ 
The (a) dense $C_{4 \times 1}^1$ and (b) sparse $C_{4 \times 1}^2$ chain vortex configurations have both (c) the topologically same effective Majorana model description. Likewise, (d) and (e) are six vortex ladder configurations $C_{3 \times 2}^1$ and $C_{3 \times 2}^2$, respectively, both described by the same effective model shown in (f). The solid, dashed and dotted lines denote $t_1$, $t_2$ and $t_{\sqrt{3}}$ interactions, respectively. To evaluate the flux per plaquette, we use counter-clockwise convention to assign the overall $i$ ($-i$) factor per link when a Majorana fermion hops along (against) the shown orientation. }
\label{effmodel}	
\end{figure}

The degeneracy lifting due to tunneling Majorana fermions suggests that this energy band can be modelled by Majorana fermions tunneling on a lattice defined by the vortex locations \cite{Grosfeld}. Such a model possesses two properties that we have already observed on the honeycomb lattice model: it has particle-hole symmetry giving rise to double spectrum and it exhibits exact zero energy states for an odd number of sites. The most general Hamiltonian one can write is
\be \label{Heff}
	H = i \sum_l \sum_{|i-j|=l} t_l s_{ij}^l \gamma_i \gamma_j,
\ee
where $t_l$ are tunneling amplitudes between sites of separation $l$, $s_{ij}^l$ are local $Z_2$ gauge degrees of freedom on link $(ij)$ and $\gamma_i$ are Majorana fermions satisfying $\gamma_i\dagger = \gamma_i$ and $\{ \gamma_i, \gamma_j \}= 2\delta_{ij}$. The Hamiltonian can be defined on arbitrary lattice geometries, which in our case is fixed by the positioning of the vortices. As the interactions in the honeycomb lattice are exponentially suppressed, we simplify the effective model by considering only the three shortest range interactions of range $l=1,\sqrt{3},2$. We refer to them as $t_1$, $t_{\sqrt{3}}$ and $t_2$ interactions, respectively, that couple sites as illustrated in Figures \ref{effmodel}(c) and \ref{effmodel}(f). The configurations $s^l = \{ s_{ij}^l \}$ are called gauge degrees of freedom, because the spectrum does not depend on them explicitly. Instead, the model depends on the introduced effective flux, which we define on plaquette $(ijk)$, $i$, $j$ and $k$ being the three corners of a triangular plaquette enumerated counter-clockwise, by
\be \label{flux}
  \Phi_{ijk} = -i \ln ( i s_{ij} s_{jk} s_{ki} ).
\ee
This evaluates always to either $\pm \frac{\pi}{2}$. The set of fluxes on all plaquettes is referred to as a \emph{flux sector}. 

The free parameters of the model are the couplings $t_l$ and the flux sectors. Our objective is to study how they are to be fixed such that the spectrum of the Majorana model will approximate the spectrum of the fusion modes. To compare quantitatively the spectrum $\bar{\mathcal{E}_n}= (-E_{n/2},-E_{n/2-1}, \ldots,E_{n/2-1},E_{n/2})$ of the fusion modes from an $n$-vortex system to the spectrum $\bar{\epsilon}_n= (-\epsilon_{n/2},-\epsilon_{n/2-1}, \ldots,\epsilon_{n/2-1},\epsilon_{n/2})$ of a corresponding $n$-site effective model, we introduce the deviation measure
\be \label{F}
 F\left[\bar{\mathcal{E}}_n,\bar{\epsilon}_n\right] = \frac{1}{n} |\bar{\mathcal{E}}_n-\bar{\epsilon}_n|.
\ee
For $F=0$ the spectra match exactly. To systematically study the many vortex systems, we restrict to considering homogenous vortex configurations, denoted by $C_{N_x \times N_y}^d$, consisting of $N_x$ vortices in direction $\mathbf{n}_x$ and $N_y$ vortices in direction $\mathbf{n}_y$, all separated by at least $d$ links. In particular, we will consider chain ($N_y=1$) and ladder ($N_y=2$) configurations that are illustrated in Figures \ref{effmodel}(a) and \ref{effmodel}(b), and Figures \ref{effmodel}(d) and \ref{effmodel}(e), respectively. Configurations that differ only by the sparsity $d$ have topologically the same effective description. For instance, all four vortex chains are described by an effective Majorana model living on the four site lattice shown in Figure \ref{effmodel}(c). 

Our strategy for studying the many vortex degeneracy lifting is as follows. We set again globally $J_x=J_y=J_z=1$, but as before leave the effective magnetic field strength $K$ as a free parameter. Considering then vortex chain and ladder configurations, we find the fusion mode spectra $\bar{\mathcal{E}}(K)$ for the range $0 \leq K \leq 0.3$. Since the fusion mode energy $\mathcal{E}^{d_s}$ in two vortex systems can be understood as arising due to tunneling between vortex cores, its magnitude at a fixed vortex separation $d_s=l$ should correlate with the tunneling amplitudes between sites of separation $l$. We assume the simplest possible correlation and use the ansatz
\be \label{tl}
t_l(K) = \mathcal{E}^{dl}(K),
\ee
for the effective model couplings corresponding to $C_{N_x \times N_y}^d$ vortex configuration. For the flux sectors, on the other hand, there is no obvious a priori correlation with the underlying model. We assume only that topologically equivalent effective models, i.e. ones differing only by the sparsity $d$, should be defined on the same flux sector. To find the one providing the correct effective description, we need to consider the effective spectrum $\bar{\epsilon}(K) = \bar{\epsilon}\left[t_l(K)\right]$ over all the possible different flux sectors.

\subsubsection{Vortex chains}

\begin{figure}[t]
\begin{center}
\begin{tabular}{cc}
\includegraphics*[width=7cm]{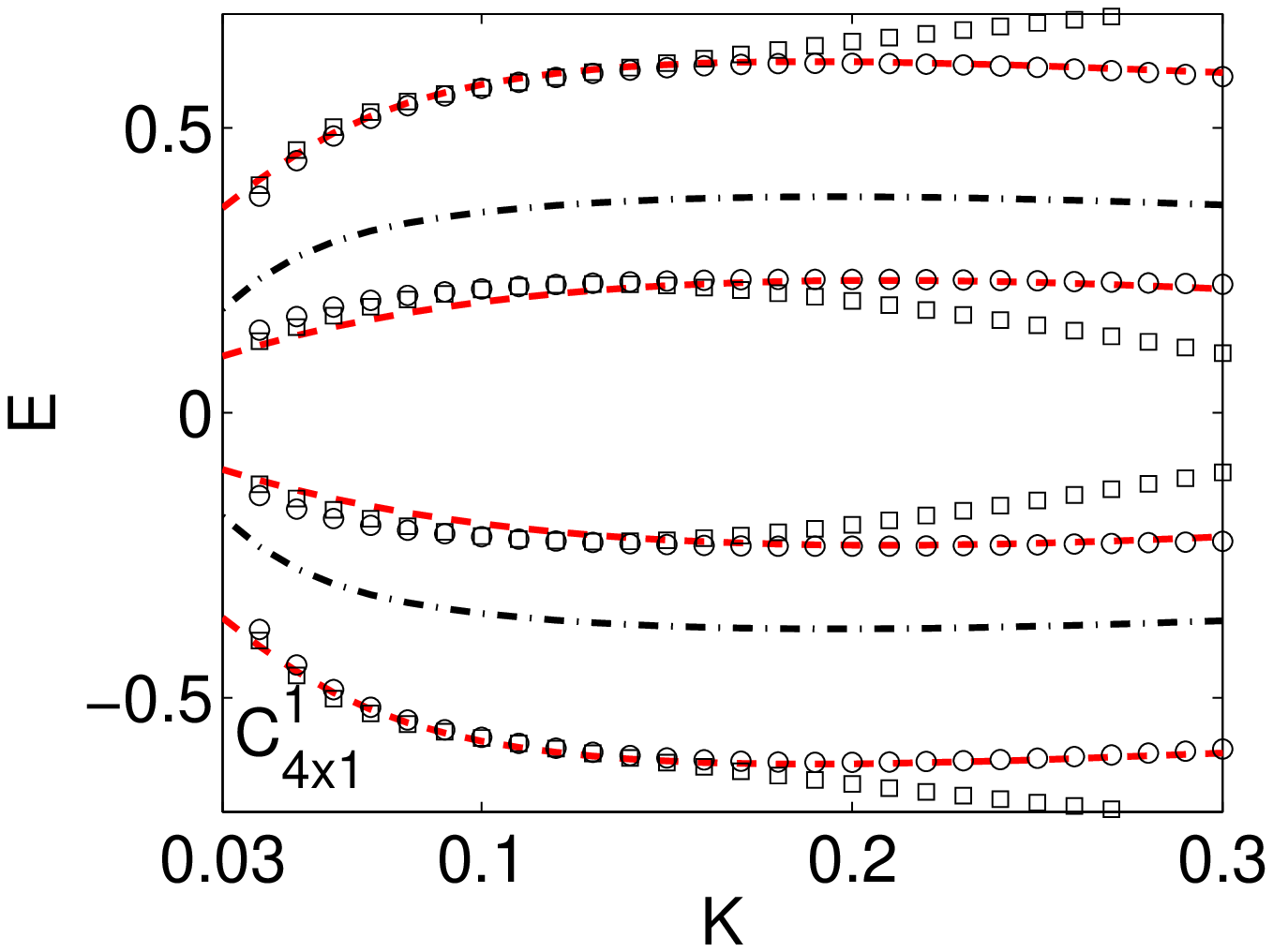} & \includegraphics*[width=7cm]{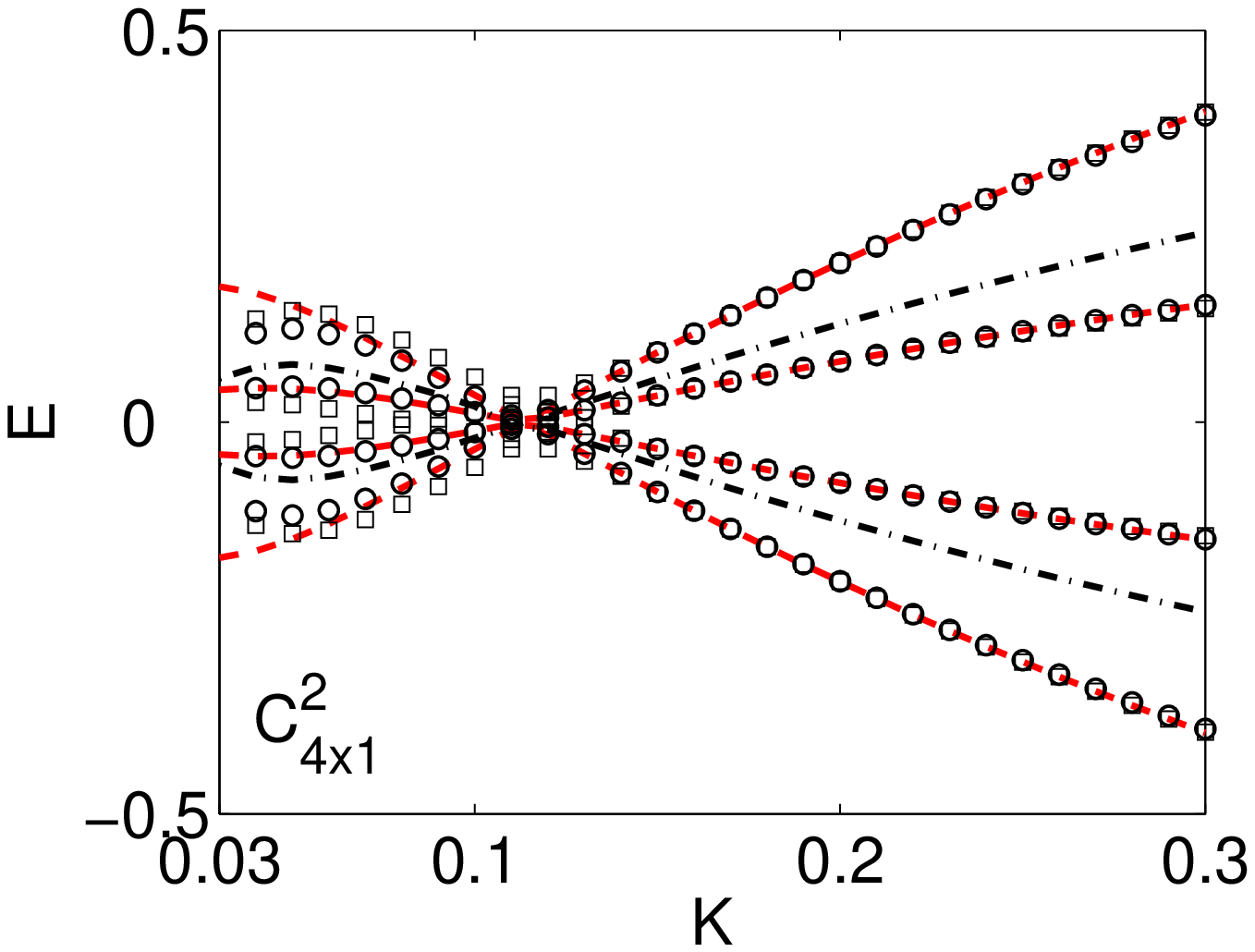} \\ (a) & (b) \\
\includegraphics*[width=7cm]{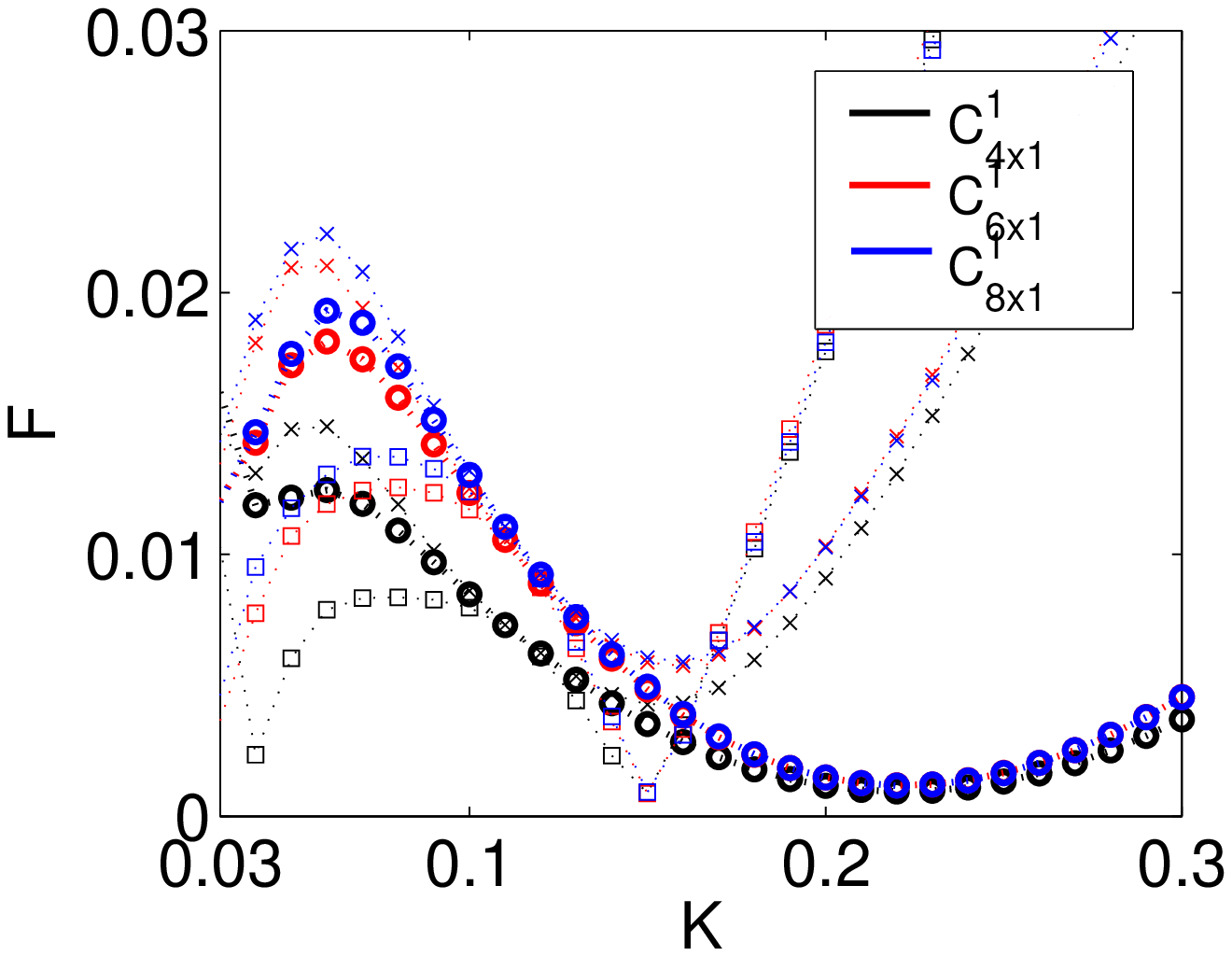} & \includegraphics*[width=7cm]{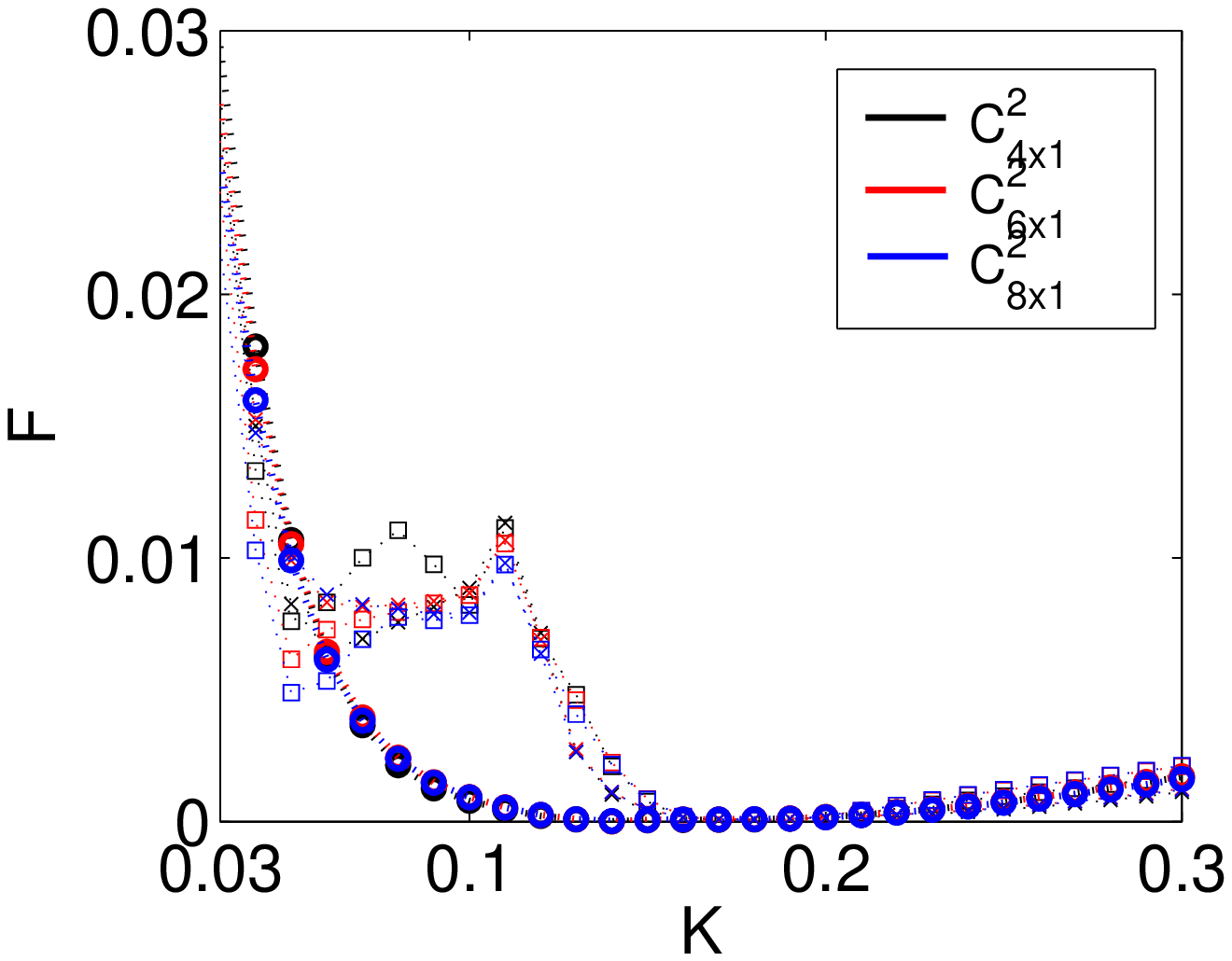} \\ (c) & (d)
\end{tabular}
\end{center}
\caption{ The fusion mode and effective model spectra for (a) the dense $C_{4 \times 1}^1$ and (b) the sparse $C_{4 \times 1}^2$ chain configurations. The red dashed lines denote fusion mode spectra $\bar{\mathcal{E}}$, the black dash dotted lines the bi-partite fusion mode energies $\mathcal{E}^{d_s}$ corresponding to $t_1$ interactions, and the squares and circles the effective model spectra with (on the alternating flux sector) and without the $t_2$ interactions, respectively. The deviations $F$ for (c) the dense chains $C_{N \times 1}^1$ and (d) the sparse chains $C_{N \times 1}^2$ for different flux sectors. The circles correspond to $t_1$ effective model only, while the crosses and squares correspond to uniform and alternating flux configurations, respectively. The approximation by the effective model compared to the pure $t_1$ model can only be improved when employing the alternating flux sector. The region $K<0.03$ is omitted due to finite size effects. }
\label{chain}
\end{figure}

Let us consider first vortex chains where the $t_{\sqrt{3}}$ interactions are naturally absent due to the vortex arrangement. Figures \ref{chain}(a) and \ref{chain}(b) show the fusion mode and the effective Majorana model spectra with nearest neighbour interactions only for the dense $C_{4 \times 1}^1$ and sparse $C_{4 \times 1}^2$ four vortex chains, respectively. In the absence of also $t_2$ interactions, there are no non-trivial plaquettes and thus the spectrum does not depend on the flux sector. We find that the $t_1$ interactions alone provide an excellent approximation, which, as quantitatively shown in Figures \ref{chain}(c) and \ref{chain}(d), corresponds to at least $F<0.02$ and gets better with increasing $K$. This is in agreement with longer range tunneling being more suppressed for larger $K$. We note also that in Figure \ref{chain}(b) the degeneracy point in the fusion mode spectrum $\bar{\mathcal{E}}(K)$ and the zero energy point of the vortex-vortex interaction induced fusion mode energy coincide exactly at $K \approx 0.12$. This strongly suggests that the nearest neighbour tunneling is primarily responsible for the degeneracy lifting in vortex chains. 

When the $t_2$ interactions are switched on, Figures \ref{chain}(c) and \ref{chain}(d) show that the approximation can be improved for $K \lesssim 0.1$ only if one chooses the fluxes alternating between $\frac{\pi}{2}$ and $-\frac{\pi}{2}$. For instance, for the $C_{4 \times 1}^1$ configuration shown in Figure \ref{effmodel}(c) one should choose $[\Phi_{123},\Phi_{243}] = \pm[\frac{\pi}{2},-\frac{\pi}{2}]$. On the other hand, for $K  \gtrsim 0.1$ we find that the inclusion of the $t_2$ interactions, regardless of the flux sector, has either neglible or detrimental effect on the approximation. The distinct behavior at the different $K$ regimes can again be traced back to the $K$ controlled localisation of the Majorana modes. For the experimentally interesting small $K$ regime the longer range interactions are more relevant and should be included in the effective model. On the other hand, for large $K$ they are either negligible due to $t_2 \ll t_1$ (sparse $d=2$ chains) or the bi-partite splitting \rf{tl} overestimates their strength due to collective effects (tight $d=1$ chains). As these observations apply also to longer vortex chains, we conclude that the $t_2$ interactions are physically relevant in the small $K$ regime of the vortex chain systems. The necessary flux sector is the one having $\pm \frac{\pi}{2}$ flux alternating on the plaquettes consisting of two $t_1$-links and a single $t_2$-link.

\subsubsection{Vortex ladders}

\begin{figure}[t]
\begin{center}
\begin{tabular}{cc}
\includegraphics*[width=7cm]{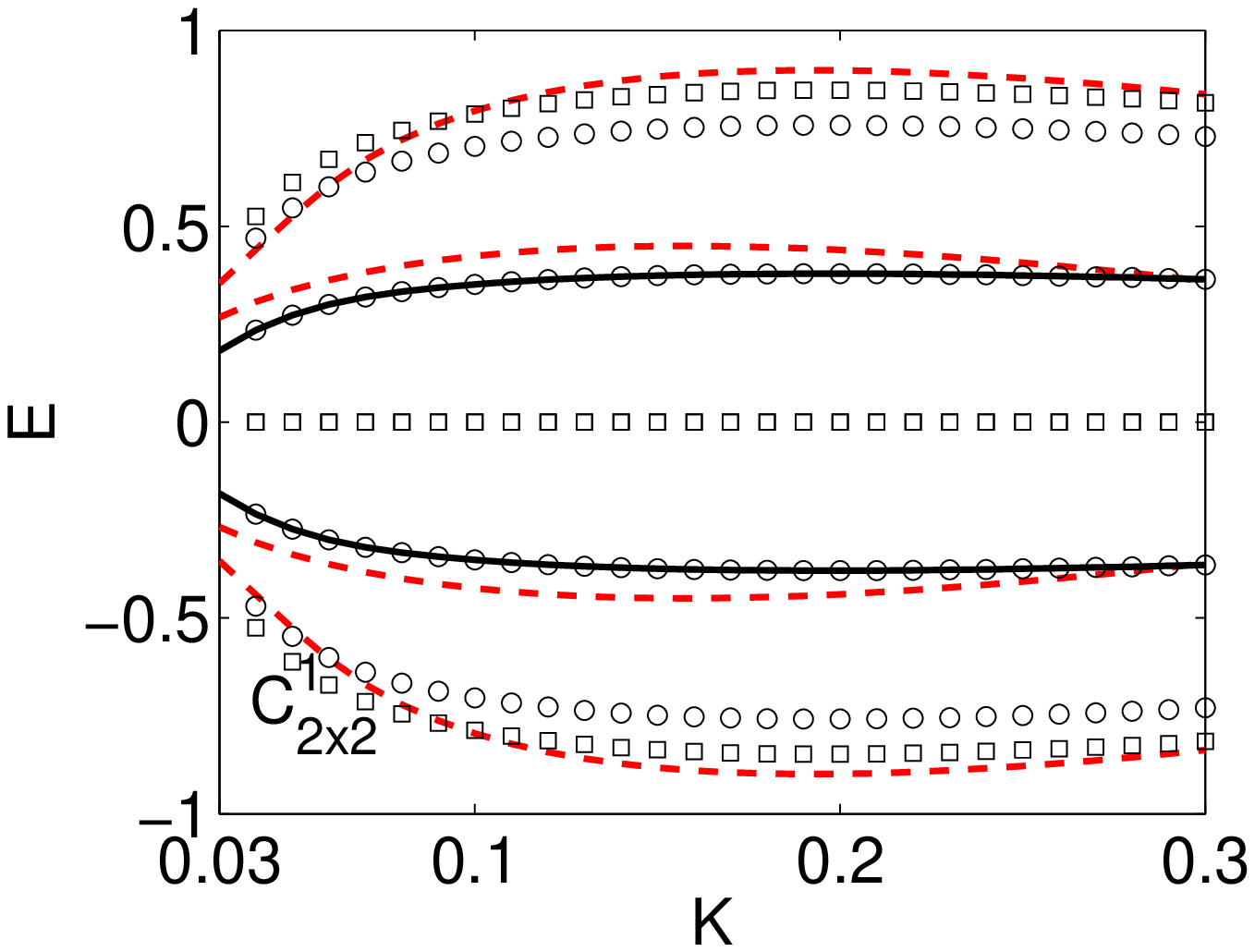} & \includegraphics*[width=7cm]{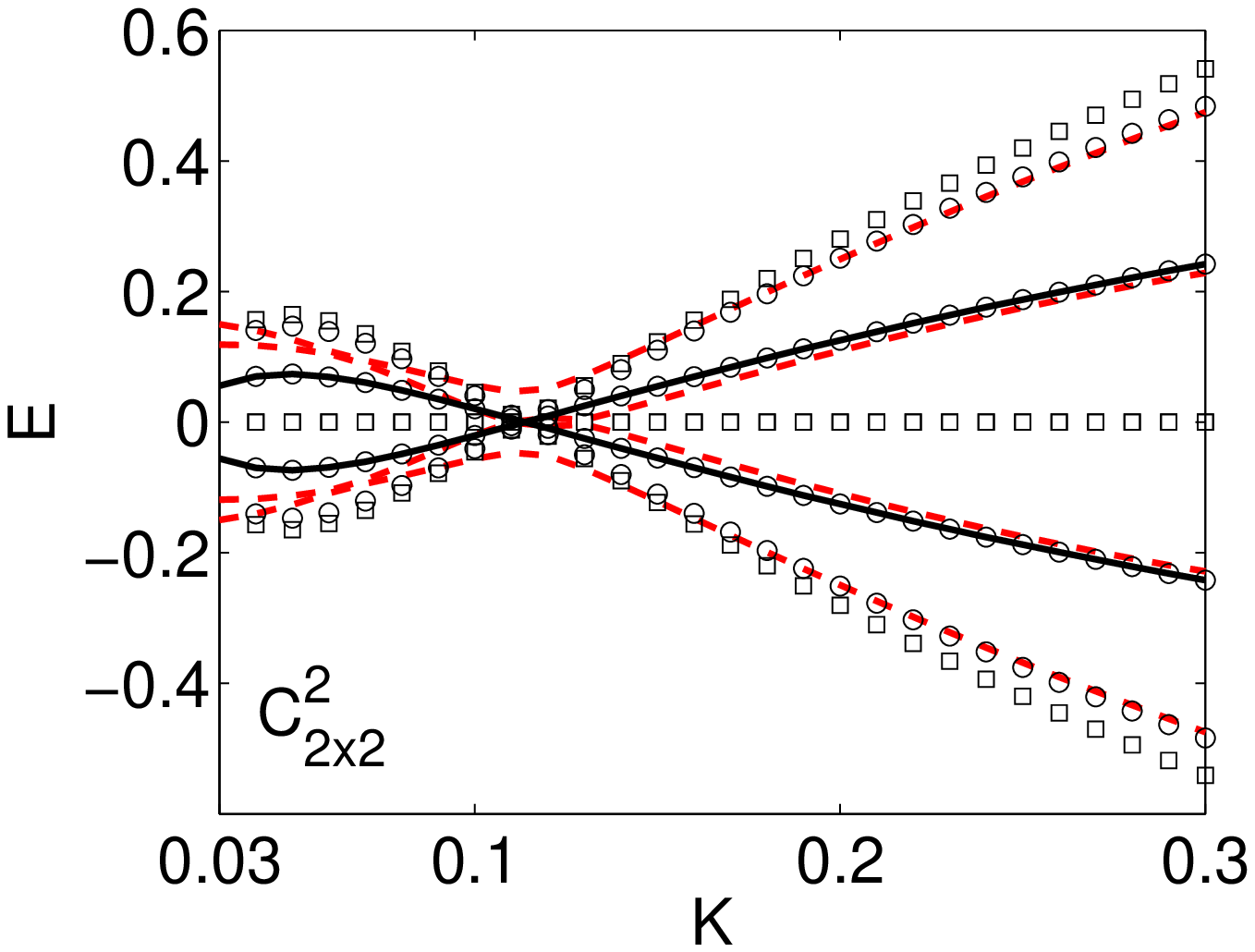} \\ (a) & (b) \\ 
\includegraphics*[width=7cm]{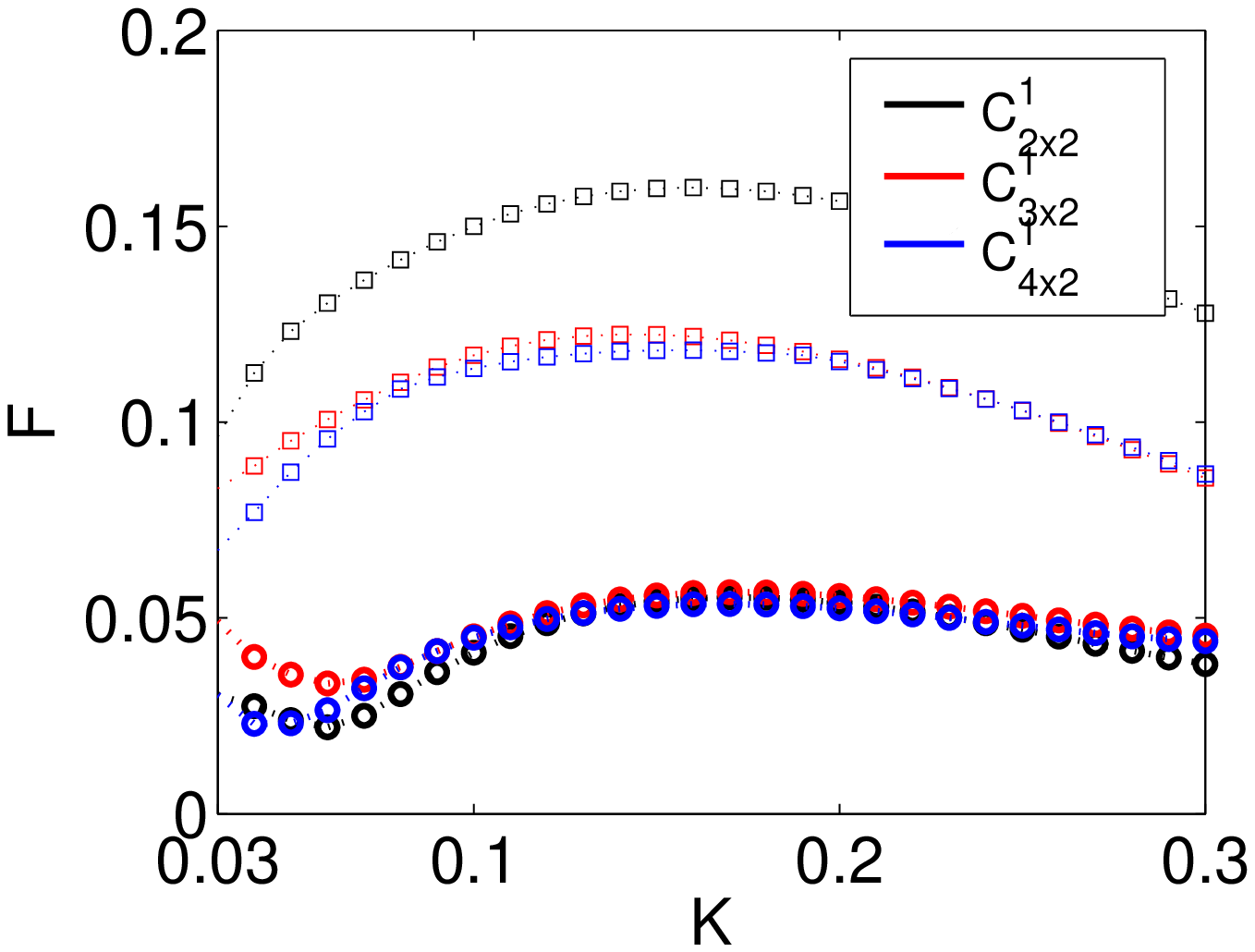} & \includegraphics*[width=7cm]{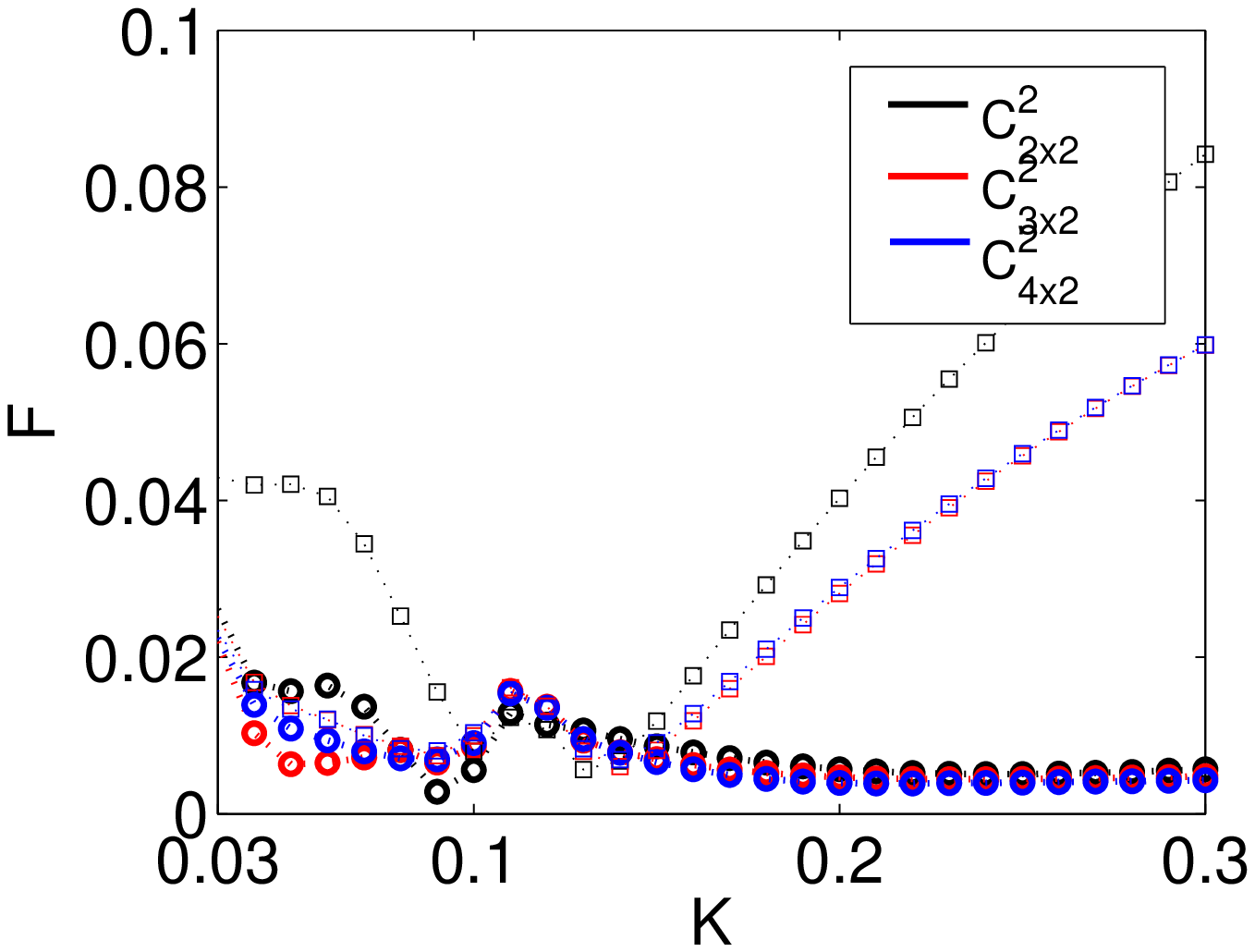} \\ (c) & (d)
\end{tabular}
\end{center}
\caption{ The fusion mode and effective model spectra for (a) the dense $C_{2 \times 2}^1$ and (b) the sparse $C_{2 \times 2}^2$ vortex ladder configurations.  The red dashed lines denote the fusion mode spectra $\bar{\mathcal{E}}$, the black dash dotted lines the fusion mode energies $\mathcal{E}^{d_s}$ corresponding to $t_1$ interactions, and the circles and squares the $t_1$ effective model spectra on the uniform and the alternating flux sectors, respectively. The deviations $F$ for (c) the dense $C_{N \times 2}^1$ and (d) the sparse $C_{N \times 2}^2$ ladders for different flux sectors. The circles and squares denote flux sectors as above. The region $K<0.03$ is omitted due to finite size effects. }
\label{ladder}
\end{figure}

The main difference between the vortex chains and the ladders is the additional presence of the $t_{\sqrt{3}}$ interactions. However, let us start again with the nearest neighbour effective model which has now two distinct flux sectors: the uniform sector with $\Phi_{126}=\Phi_{256}=\Phi_{235}=\Phi_{345}$ or the alternating sector with $\Phi_{126}=\Phi_{235}=-\Phi_{256}=-\Phi_{345}$ (see Figure \ref{effmodel}(f)). Figures \ref{ladder}(a) and \ref{ladder}(b) for the smallest dense and sparse ladders, respectively, show that a qualitatively good fit can be obtained in the uniform flux sector, whereas in the non-uniform sector the spectrum is gapless. This is reflected in the deviation measures, shown in Figures \ref{ladder}(c) and \ref{ladder}(d), where the uniform flux sector clearly provides a better approximation than the non-uniform one. Therefore, we conclude that having a $\frac{\pi}{2}$-flux on all plaquettes consisting of $t_1$ links only is another necessary condition for the effective Majorana model to approximate the full model.

In general the approximation by the nearest neighbour model is not as good for the vortex ladders as it was for the vortex chains. However, this error is in general systematic and can be compensated by scaling the tunneling amplitudes as $t_l = a \mathcal{E}^{dl}$ for some $K$ indepedent constant $a$. While the systematic study of such dressing effects is beyond the present work, one can imagine them occurring due to quantum interference of many Majorana wave functions. Such effects are also likely to play a role when the $t_{\sqrt{3}}$ and $t_2$ interactions are switched on. In this case the flux needs to be defined also on plaquettes consisting of two $t_1$-links and a single $t_{\sqrt{3}}$-link, and on plaquettes with one link of each type. For instance, in Figure \ref{effmodel}(f) these correspond to the fluxes $[\Phi_{156},\Phi_{125},\Phi_{245},\Phi_{234},\Phi_{356},\Phi_{362}]$ and $[\Phi_{136},\Phi_{135}, \Phi_{462},\Phi_{463}]$. We find that the approximation can in general be improved for some $K$ ranges by introducing the longer range interactions, but unlike in the vortex chains we find no single flux sector that would systematically improve it in the physically relevant small $K$ regime. We attribute this to the studied vortex arrays being too small for such genuinely two dimensional effects to become transparent. We leave their systematic study for future work \cite{Lahtinen_inprep}.

\subsection{Effective flux sectors and the underlying $\pi$-flux vortices}

\begin{figure}[t]
\begin{center}
\begin{tabular}{ccc}
\includegraphics*[width=3.7cm]{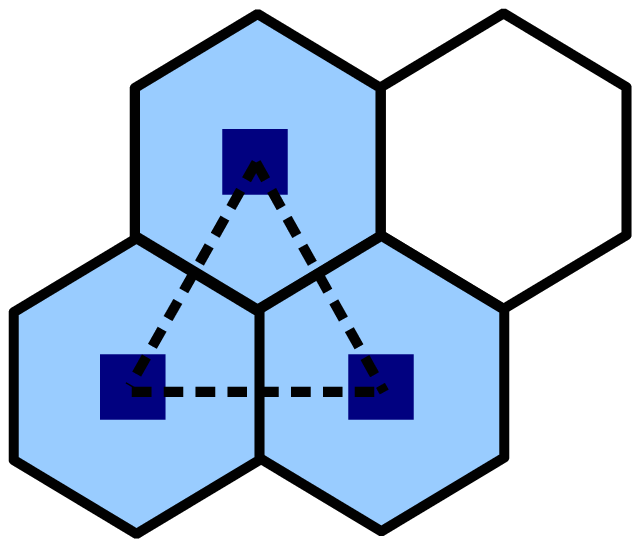} & \includegraphics*[width=3.7cm]{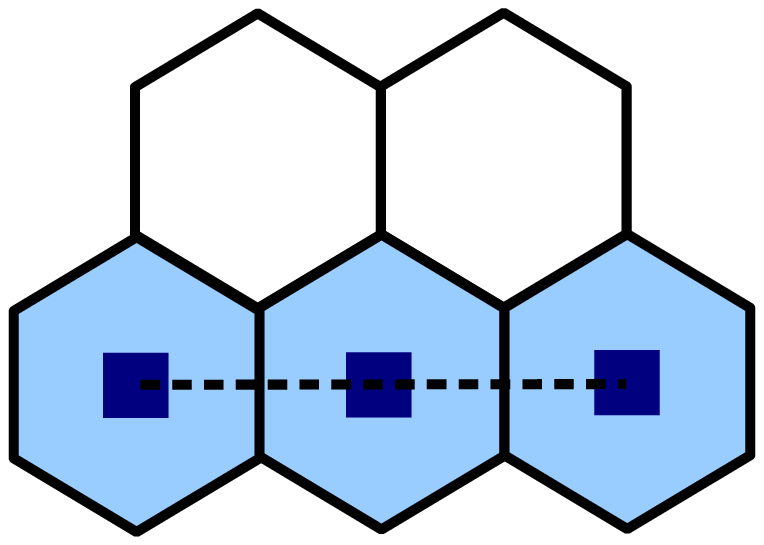} & \includegraphics*[width=3.7cm]{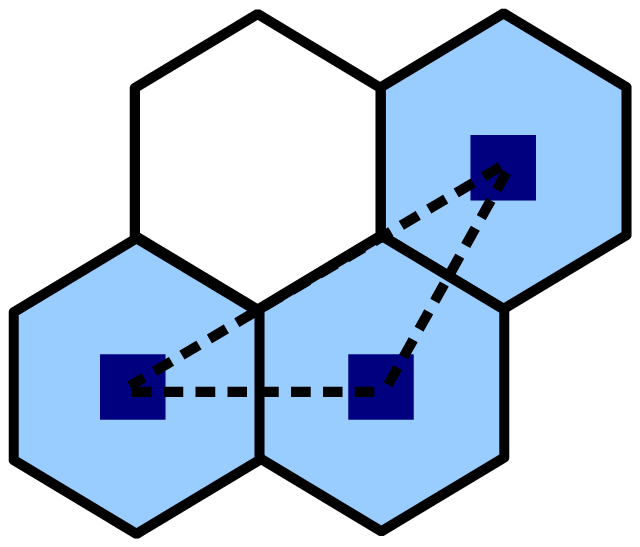} \\ (a) & (b) & (c) \end{tabular}
\end{center}
\caption{ Flux on the effective model arises from the $\pi$-flux vortices of the underlying model. When the flux of a vortex is assumed to be uniformly spread across the plaquette, the flux through the shown effective model plaquettes is given by (a) $\Phi = \frac{\pi}{6}+\frac{\pi}{6}+\frac{\pi}{6} = \frac{\pi}{2}$, (c) $\Phi = 0$ (c) $\Phi = \frac{\pi}{12}+\frac{\pi}{12}+\frac{\pi}{3} = \frac{\pi}{2}$. }
\label{fluxgen}
\end{figure}

We discovered that having a $\frac{\pi}{2}$-flux on every plaquette consisting of only $t_1$-links is a necessary condition for the effective Majorana model to approximate the full one. Also, in the presence of $t_2$ interactions, the approximation for vortex chains could be improved only when one imposed an alternating $\pm \frac{\pi}{2}$-flux pattern on the plaquettes consisting of two $t_1$-links and a single $t_2$-link. We postulate that these conditions are connected to the underlying honeycomb lattice model as follows. The anyonic vortices there are $\pi$-flux vortices. Let us assume the flux to be uniformly spread across the hexagonal plaquette a vortex occupies and the Majorana mode to be bound exactly at its center. Then, as illustrated in Figure \ref{fluxgen}(a), a triangular dual lattice plaquette consisting of three $t_1$-links will enclose exactly $\frac{\pi}{2}$-flux arising collectively from the three vortices. On the other hand, the plaquettes consisting of a single $t_2$ link and two $t_1$ links, as illustrated in Figure \ref{fluxgen}(b), span no area and should thus have no flux on them. While trivial flux can not directly imposed on the Majorana model, we interpret the alternating $\pm \frac{\pi}{2}$-flux sector effectively providing it.

We postulate that when writing down a general effective Majorana model for interacting $\pi$-flux vortices, the flux on every possible effective model plaquette should be chosen to be consistent with the enclosed vortex flux from the underlying microscopic model. For instance, as illustrated in Figure \ref{fluxgen}(c), the plaquettes consisting of a single $t_{\sqrt{3}}$-link and two $t_1$-links should have also $\frac{\pi}{2}$-flux. While the concept of enclosed vortex flux is equivalent of the net phase in vortex arrays \cite{Grosfeld}, we emphasize the conceptual simplicity of our prescription in relating the microscopic model and the necessary conditions it imposes on the effective model. The verification of this picture and the development of a thermodynamic Majorana model is a subject of future research \cite{Lahtinen_inprep}.

\section{Conclusions}

We have studied the interacting non-Abelian Ising anyons emerging from Kitaev's honeycomb lattice model. By simulating continuous vortex transport \cite{Lahtinen09}, we uncovered the characteristic oscillations in the bi-partite degeneracy lifting and characterized their physical parameter dependence. While the oscillations had been discovered earlier for both $p$-wave superconductors \cite{Cheng09} and the Moore-Read state \cite{Baraban09}, these calculations involve mean fields and trial wave functions, respectively. Our results demonstrate the oscillations in an actual microscopic model. Due to algebraic similarity, we expect them to be present also in the variations of the honeycomb lattice model \cite{Yao07,Kells10-2,Yang07,Yao09,deGottardi11}. Furthermore, we extended the results of \cite{Lahtinen08} by demonstrating unambiguously the fusion degrees of freedom and by obtaining the energy gaps characterizing the stability of the topological low energy theory. 

In the second part we studied degeneracy lifting in interacting many vortex systems. Employing the picture of localized Majorana modes, we wrote down a Majorana fermion model on a finite lattice whose sites coincide with the vortex locations. By comparing its spectrum to that of the fusion modes of a corresponding vortex configuration, we were able to find necessary conditions for the spectra to match. This amounted to finding relevant flux sectors of the effective model, which we interpreted as arising from the $\pi$-flux vortices of the underlying full model. These results are consistent with and extend the earlier mean-field studies in the context of $p$-wave superconductors \cite{Grosfeld}. Further, we found that the energy splitting due to vortex-vortex interactions could be employed to high accuracy as the effective tunneling amplitude of the Majorana fermions. This confirms that bi-partite tunneling is predominantly responsible for the degeneracy lifting also in many vortex systems. If many body effects are present, they can be included as dressed or longer range couplings. Our results pave the way for constructing an effective model for interacting anyonic $\pi$-flux vortices at the thermodynamic limit \cite{Lahtinen_inprep}. Such model can be employed to understand phase transitions due to vortex lattices, \cite{Lahtinen10}, and test the general theory of anyon-anyon interaction driven phase transitions in the context of the microscopic honeycomb lattice model \cite{Ludwig11,Lahtinen_inprep}.

\section*{Acknowledgements}

VL would like to thank Jiannis Pachos, Simon Trebst and Andreas Ludwig for inspiring discussions as well as Station Q and ETH Zurich for hospitality. This work is supported by EPSRC and the Finnish Academy of Science.

\end{document}